\documentclass[letterpaper,10pt,twocolumn]{IEEEtran}
\usepackage{amssymb,amsmath,epsfig,amsthm}
\usepackage[latin5]{inputenc}
\usepackage{graphicx,times,latexsym}
\usepackage{subcaption}
\usepackage{psfrag,cite}
\usepackage{setspace}
\usepackage{algorithmic, algorithm}
\usepackage{xcolor}

\usepackage{hyperref}

\usepackage{wasysym}

\usepackage{mathtools}
\mathtoolsset{showonlyrefs=true}

\usepackage{bbm}

\newtheorem{theorem}{Theorem}
\newtheorem{lemma}{Lemma}

\newtheorem{corollary}{Corollary}

\theoremstyle{remark}
\newtheorem{remark}{Remark}

\theoremstyle{definition}
\newtheorem{defn}{Definition}
\newtheorem{exmp}{Example}

\newcommand{\blue}[1]{{\color{black} #1}}

\usepackage{amssymb,amsmath,epsfig}
\usepackage[latin5]{inputenc}
\usepackage{graphicx,times,latexsym}
\usepackage{subcaption}
\usepackage{setspace}
\usepackage{algorithmic, algorithm}
\usepackage{color}
\usepackage{bbm}

\newcommand{\oprocendsymbol}{\hbox{$\bullet$}}
\newcommand{\oprocend}{\relax\ifmmode\else\unskip\hfill\fi\oprocendsymbol}

\providecommand{\norm}[1]{\left\| #1 \right\|}
\providecommand{\Pdec}{P_{\mathrm{Dec}}}
\providecommand{\Pdet}{P_\mathrm{Det}}
\providecommand{\Pfa}{P_\mathrm{FA}}

\providecommand{\figref}{Fig.\,\ref}
\providecommand{\secref}{Sec.\,\ref}
\providecommand{\thmref}{Thm.\,\ref}
\providecommand{\lemref}{Lem.\,\ref}
\providecommand{\colref}{Corol.\,\ref}
\newcommand{\nats}{\mathbb{N}}
\newcommand{\real}{\mathbb{R}}
\providecommand{\hA}{\hat{A}}
\providecommand{\bhA}{\hat{\textbf{A}}}
\providecommand{\hF}{\hat{F}}

\providecommand{\htheta}{\hat{\Theta}}
\providecommand{\tW}{{\tilde{W}}}
\providecommand{\btW}{{\tilde{\textbf{w}}}}
\providecommand{\tX}{{\tilde{X}}}
\providecommand{\tZ}{{\tilde{Z}}}
\providecommand{\oJ}{\bar{J}}
\providecommand{\oU}{\bar{U}}
\providecommand{\oX}{\bar{X}}
\providecommand{\oZ}{\bar{Z}}
\providecommand{\iid}{\mathrm{i.i.d.}}

\usepackage{bm}
\providecommand{\bU}{\textbf{u}}
\providecommand{\bV}{\textbf{v}}
\providecommand{\bW}{\textbf{w}}
\providecommand{\bX}{\textbf{x}}
\providecommand{\bY}{\textbf{y}}
\providecommand{\bDelta}{\bm{\Delta}}
\providecommand{\bSigma}{\mathbf{\Sigma}}
\providecommand{\tbW}{\tilde{\bW}}
\providecommand{\cF}{\mathcal{F}}
\providecommand{\cK}{\mathcal{K}}

\providecommand{\bA}{\mathbf{A}}
\providecommand{\bB}{\mathbf{B}}
\providecommand{\bG}{\mathbf{G}}
\providecommand{\cU}{\mathcal{U}}

\providecommand{\Id}{\mathtt{I}}
\providecommand{\bzero}{\mathbf{0}}

\providecommand{\E}[1]{\mathbb{E} \left[ #1 \right]}
\providecommand{\CE}[2]{\mathbb{E} \left[ #1 \middle| #2 \right]}

\providecommand{\Esub}[2]{\mathbb{E}_{\mathbb{P}_{#1}} \left[ #2 \right]}
\providecommand{\Ep}[1]{\mathbb{E}_{\mathbb{P}_a} \left[ #1 \right]}
\providecommand{\Epi}[1]{\mathbb{E}_{\mathbb{P}} \left[ #1 \right]}
\providecommand{\Var}[1]{\mathrm{Var} \left[ #1 \right]}

\providecommand{\MI}[2]{I \left( #1 ; #2 \right)}
\providecommand{\CMI}[3]{I \left( #1 ; #2 \middle| #3 \right)}
\providecommand{\KL}[2]{D \left( #1 \middle\| #2 \right)}
\providecommand{\CKL}[3]{D \left( #1 \middle\| #2 \middle| #3 \right)}
\newcommand{\argminA}{\mathop{\mathrm{argmin}}}  

\providecommand{\inner}[2]{\left\langle #1, #2  \right\rangle}

\providecommand{\appref}[1]{App.~\ref{#1}}

\newcommand{\VersionLength}{long}

\providecommand{\ver}{\ifthenelse{\equal{\VersionLength}{long}}}
\providecommand{\nver}{\ifthenelse{\equal{\VersionLength}{short}}}

\providecommand{\second}[2]{#2}

\newcommand{\rem}[1]{}

\begin{document}
\title{Learning-based Attacks in Cyber-Physical Systems}
\author{Mohammad Javad Khojasteh, Anatoly Khina, Massimo Franceschetti, and Tara Javidi 
	\thanks{The material in this paper was presented in part at
the 8th IFAC Workshop on Distributed Estimation and Control in Networked Systems, 2019~\cite{khojasteh2019authentication}.
	This research was partially supported by NSF awards CNS-1446891 and ECCS-1917177.
    This work has received funding from the European Union's Horizon 2020 research and innovation programme under the Marie Sk\l odowska-Curie grant agreement No 708932.}
    \thanks{
    M.~J.~Khojasteh is with
    the Center for Autonomous Systems and Technologies (CAST), California Institute of
    Technology, Pasadena, CA
91125, USA. Some of this work was performed while at University of California, San Diego (e-mail: \texttt{mjkhojas@caltech.edu}).}
    \thanks{A.~Khina is with the School of Electrical Engineering, Tel Aviv University, Tel Aviv, Israel~6997801 (e-mail: \texttt{anatolyk@eng.tau.ac.il}).}
    \thanks{M.~Franceschetti, and T.~Javidi are with the Department of Electrical and Computer Engineering, University of California, San Diego, La Jolla, CA~92093, USA (e-mails: \texttt{\{mfranceschetti, tjavidi\}@eng.ucsd.edu}).}
} 

\maketitle
\IEEEpeerreviewmaketitle
 \begin{abstract}                
 We introduce the problem of learning-based attacks in a simple abstraction of cyber-physical systems---
 the case of a 
 discrete-time, linear, time-invariant plant that may be subject to an attack that overrides the sensor readings and the controller actions. The attacker attempts to learn the dynamics of the plant and subsequently overrides the controller's actuation signal, 
 to destroy
 the plant without being detected. 
 The attacker can feed fictitious sensor readings to the controller using its estimate of the plant dynamics and mimic the legitimate plant operation. 
 The controller, on the other hand, is constantly on the lookout
 for an attack;
 once the controller detects an attack,
 it 
 immediately shuts the plant off. 
 In the case of scalar plants, we derive an upper bound on the attacker's deception probability for \textit{any measurable} control policy when the attacker uses \textit{an arbitrary} learning algorithm to estimate the system dynamics.
 We then derive lower  bounds for the attacker's deception probability for both scalar and vector plants by assuming an    authentication test that 
 inspects the empirical variance of the system disturbance. 
 We  also show how the controller can improve the security of the system   by superimposing a carefully crafted \textit{privacy-enhancing signal} on top of the ``nominal~control~policy.''
 Finally, 
 for nonlinear scalar dynamics that belong to the Reproducing Kernel Hilbert Space (RKHS), we investigate the performance of attacks based on nonlinear Gaussian-processes (GP) learning algorithms.
%
%

\end{abstract}

\allowdisplaybreaks

\section{Introduction}

Recent technological advances in wireless communications and 
computation, and their integration into networked control and cyber-physical systems (CPS),
open the door to a myriad of new and exciting applications, including cloud robotics and automation~\cite{kehoe2015survey}.
However, the distributed nature of CPS
is often a source of vulnerability. 
Security breaches in CPS can have catastrophic consequences 
ranging from hampering the economy by obtaining financial gain, to hijacking autonomous vehicles and drones,   to
terrorism by manipulating life-critical infrastructures~\cite{urbina2016limiting,dibaji2019systems,jamei2016micro}.
Real-world instances of security breaches in CPS, that were discovered and made   public, include 
the revenge sewage attack in Maroochy Shire, Australia; 
the Ukraine power grid cyber-attack; 
the German steel mill cyber-attack; the Davis-Besse nuclear power plant 
attack in Ohio, USA;
and the Iranian uranium-enrichment facility attack via the Stuxnet malware~\cite{sandberg2015cyberphysical}.
Studying and preventing such security breaches via control-theoretic methods has received 
a great deal of 
attention in recent years \cite{bai2017data,dolk2017event,shoukry2018smt,chen2018cyber,shi2018causality,dibaji2018secure,tunga2018tuning,niu2018minimum,chong2019tutorial,tomic2018design,ding2018attacks,teixeira2015secure,xue2012security,cetinkaya2017networked,brown2018security,law2014security,shames2017security,hashemi2018comparison}.

An important and widely studied class of attacks in CPS is based on the ``man-in-the-middle"  (MITM) paradigm~\cite{smith2011decoupled}:
an attacker overrides the  sensor signals transmitted from the physical plant to the controller with fake signals that mimic stable and safe operation. At the same time, the attacker also overrides the control signals 
with malicious inputs to push the plant toward a catastrophic trajectory. 
It follows that CPS must constantly monitor the plant outputs and look for anomalies in the fake sensor signals
to detect
such attacks. The attacker, on the other hand, aims to 
generate fake sensor readings in a way
that would be indistinguishable from the  legitimate ones.  

The MITM attack has been  extensively  studied in two special cases~\cite{MoSinopoli:Magzine,zhu2014performance,miao2013stochastic,Kumar:DynamicWatermarki,smith2011decoupled}. The first case  is the \textit{replay attack}, in which the attacker observes and records the legitimate system behavior for a given  time window and then replays this recording periodically at the controller's input \cite{MoSinopoli:Magzine,zhu2014performance,miao2013stochastic}. The second  case is the \emph{statistical-duplicate attack}, which assumes that the attacker has acquired complete knowledge of the dynamics and parameters of the system, and can  construct arbitrarily long fictitious sensor readings  that are statistically identical to the actual signals\blue{~\cite{Kumar:DynamicWatermarki,smith2011decoupled,hespanhol2018statistical}}. The replay attack assumes no knowledge of the system parameters---and as a consequence, it is relatively easy to detect. An effective way to counter the replay attack consists of superimposing a random watermark signal, unknown to the attacker,~on top of the control signal~\cite{fang2017cost,hespanhol2018statistical,hosseini2016designing,ferdowsi2018deep,liu2018line}. 
 The statistical-duplicate attack assumes full knowledge of the system dynamics---and as a consequence, it requires a more sophisticated detection procedure, as well as additional assumptions on the attacker or controller behavior to ensure it can be detected.  To combat the attacker's full  knowledge, the controller may adopt  \emph{moving target}~\cite{weerakkody2015detecting,kanellopoulos2019moving,zhang2019analysis,griffioen2019optimal} or \emph{baiting}~\cite{ flamholz2019baiting,hoehn2016detection}~techniques. Alternatively,  the controller may introduce  private randomness in the control input  using  
\textit{watermarking} \cite{Kumar:DynamicWatermarki}.
In this scenario, a vital assumption is made: although the attacker observes the true sensor readings,~it is barred from observing the control actions,
as otherwise, it~would be omniscient and
undetectable.


Our \textbf{contributions} are as follows. First, 
we observe that  in many practical situations  the attacker does not have full knowledge of the system  and cannot simulate a statistically indistinguishable copy of the system. On the other hand, the attacker can carry out more sophisticated attacks than simply replaying previous sensor readings, by attempting to ``learn'' the system dynamics from the observations. 
For this reason, we study \emph{learning-based attacks}, in which the attacker attempts to learn a model of the plant dynamics, and show that they can outperform replay attacks on linear systems by providing a lower bound on  the attacker's deception probability   using a simple learning algorithm. 
Secondly, we derive a converse bound  on the  attacker's  deception probability in the special case of scalar systems. This holds for \textit{any} (measurable) control policy,
and for \textit{any} learning algorithm that may be used by the attacker to estimate the dynamics of the plant. \blue{These contributions regard the possibility of performing learning-based attacks.
Another contribution regards the way to defend the system against these attacks.
}
For any learning algorithm utilized by the attacker to estimate the dynamics of the plant, we show that adding a proper \textit{privacy-enhancing signal} to 
the ``nominal control policy''
can lower the deception probability. Finally, we offer a treatment for nonlinear scalar dynamics that belong to a Reproducing Kernel Hilbert Space (RKHS), by studying the performance of a nonlinear attack based on machine-learning GP algorithms.

Throughout the paper, we assume that the attacker has  full access to both sensor and  control signals. The controller, on the other hand, has perfect knowledge of 
the system dynamics and tries to discover the attack from the  observations that are maliciously injected by the attacker. 
This assumed information-pattern \textit{imbalance} between the controller and the attacker is justified since the controller is \textit{tuned} in much longer than the attacker and thus has knowledge of the system dynamics to a far greater precision than the attacker. On the other hand, the attacker can completely hijack the sensor and control signals that travel through a communication network that has been compromised.  \blue{Previous watermarking techniques~\cite{MoSinopoli:Magzine,hespanhol2018statistical,Kumar:DynamicWatermarki} are only effective at securing the system if the attacker has no access to the control signals, which is not the case here. On the other hand, since in our case, the attacker does not have full knowledge of the system dynamics, our privacy-enhancing signal is used to hamper the learning process of the attacker during the learning phase, rather than providing a unique signature to the control signal as in the case of watermarking.}

Since in our setting the success or failure of the attacker
is dictated by   its learning capabilities, our work is also related to   recent studies in learning-based control~\cite{Dean2019,deisenroth2011pilco,gal2016improving,rantzer2018concentration,tu2017leastrecht,sarkar2019near,berkenkamp2017safe,fisac2018general,khojasteh2019probabilistic22,cheng2020safe}. \blue{In contrast to these works, 
where tools developed in machine learning are used to design controllers in the presence of uncertainty, our work  assumes a   setting in which the controller has perfect knowledge of the system dynamics and tries to discover a possible attack from the observations. At the same time,   the attacker aims to learn the system dynamics, to construct a carefully crafted fictitious sensor reading signal to fool the controller. Thus, the security guarantees in this work are achieved by analyzing the performance and limitations of learning algorithms.}

Learning-based attacks are also related to the Known-Plaintext Attacks (KPA), introduced in~\cite{yuan2015security}, in linear systems with linear controllers. Using pole--zero analysis from classical system identification, \cite{yuan2015security} investigates necessary and sufficient conditions for which the system is 
identifiable by  an attacker  and, as a result, vulnerable against KPA. 
To combat KPA, \cite{yuan2015security} utilizes low-rank linear controllers that trade control performance for security.

The outline of the rest of the paper is as follows.
The notations used in this work are detailed in \secref{ss:notation}.
For ease of exposition, we start by presenting the problem for the special case of scalar linear plants
in \secref{s:model}, and present our main results for this case in \secref{Statementwff2}.
We then extend the model and treatment to vector linear  and scalar nonlinear  plants in \secref{sec:vector2} and \appref{futrelatef2}, respectively. 
We conclude the paper in \secref{sec:conc2} 
with a discussion of future directions. 


\vspace{-1\baselineskip}
\subsection{Notation}
\label{ss:notation}
\begin{figure}[t]
	\begin{subfigure}[c]{\columnwidth}
	\centering
   	\includegraphics[scale=0.34]{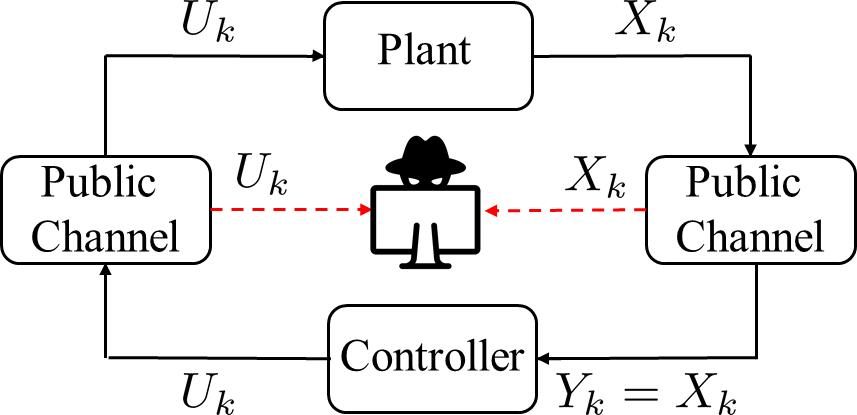}
        \caption{Learning: During this phase, the attacker eavesdrops and learns the system, without altering the input signal to the controller ($Y_k = X_k$).}
        \label{fig:sn1}
  \end{subfigure}
  
  \ \\ 
  
    \begin{subfigure}{\columnwidth}
	\centering
     	\includegraphics[scale=0.36]{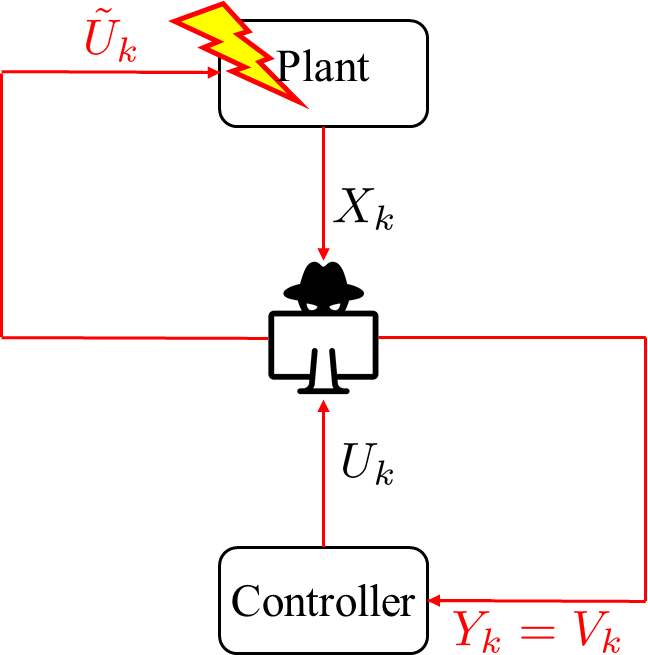}
        \caption{Hijacking: During this phase, the attacker hijacks the system and intervenes as a MITM in two places: acting as a fake plant for the controller ($Y_k=V_k$) by impersonating the legitimate sensor, and as a malicious controller ($\tilde{U}_k$) for the plant aiming to destroy the plant.}
        \label{fig:sn2}
   \end{subfigure}
   \caption{System model during learning-based attack phases.}
   \label{fig:sns}
\end{figure}

Throughout the paper, we use the following notation. We denote by $\mathbb{N}$ the set of natural numbers, and by $\real$---the set of real numbers.
All logarithms, denoted by $\log$, are base 2. 
We denote by $\norm{\cdot}$ the Euclidean norm of a vector, 
and by $\norm{\cdot}_{op}$---the 
\textit{operator}
norm induced by it when applied to a matrix.
We denote by $\dagger$ the transpose operation of a matrix.
For two real-valued functions $g$ and $h$,
$g(x) = O \left( h(x) \right)$ as $x \rightarrow x_0$
means 
$\limsup_{x \rightarrow x_0} \left| g(x) / h(x) \right| < \infty$, 
and $g(x) = o \left( h(x) \right)$ as $x \rightarrow x_0$ means 
$\lim_{x \rightarrow x_0} \left| g(x) / h(x) \right| = 0$. 
We denote by $x_i^j=(x_i,\cdots,x_j)$ the realization of 
the tuple of random variables $X_i^j=(X_i,\cdots,X_j)$ for $i, j \in \mathbb{N}, i \leq j$. Random matrices are represented by boldface capital letters (e.g.~\textbf{A}) and their realizations are represented by typewriter boldface letters (e.g.~\texttt{A}). 
$\bA \succeq \bB$ means that $\bA - \bB$ is a positive semidefinite matrix, namely $\succeq$ is the Loewner order of Hermitian matrices. $\lambda_{max}(\bA)$ denotes the largest eigenvalue of the matrix $\bA$. 
\blue{We represent the random vector with boldface small letters, and $\textbf{x}_i^j=(\textbf{x}_i,\cdots,\textbf{x}_j)$ for $i, j \in \mathbb{N}, i \leq j$.
$\mathbb{P}_\textbf{x}$} denotes the distribution of the random vector $\textbf{x}$ with respect to (w.r.t.) 
probability measure $\mathbb{P}$, 
whereas $f_\textbf{x}$ denotes its probability density function (PDF)
w.r.t.\ to the Lebesgue measure, if it has one.
An event is said to happen almost surely (a.s.) if it occurs with probability one. For real numbers $a$ and $b$, $a\ll b$ means $a$ is much less than $b$, in some numerical sense, while for probability distributions $\mathbb{P}$ and $\mathbb{Q}$, $\mathbb{P} \ll \mathbb{Q}$ means $\mathbb{P}$ is absolutely
continuous w.r.t.\ $\mathbb{Q}$. 
$d\mathbb{P}/d\mathbb{Q}$ denotes the Radon--Nikodym derivative of $\mathbb{P}$ w.r.t.\ $\mathbb{Q}$. 
The Kullback--Leibler (KL) divergence between 
probability distributions $\mathbb{P}_X$ and $\mathbb{P}_Y$
is defined as
\begin{align*}
D(\mathbb{P}_X\|\mathbb{P}_Y) \triangleq
    \begin{cases}
        \Esub{X}{\log\dfrac{d\mathbb{P}_X}{d\mathbb{P}_Y}}, & \mathbb{P}_X \ll \mathbb{P}_Y;
	 \\ \infty, & \textrm{otherwise} ,
    \end{cases}
\end{align*}
where $E_{\mathbb{P}_X}$ denotes the expectation w.r.t.\ probability measure $\mathbb{P}_X$.
The conditional KL divergence between probability distributions $\mathbb{P}_{Y|X}$ and $\mathbb{Q}_{Y|X}$ averaged over $\mathbb{P}_X$ is defined as $D \left( \mathbb{P}_{X|Y} \middle\| \mathbb{Q}_{Y|X} \middle| \mathbb{P}_{X} \right) \triangleq  \mathbb{E}_{\mathbb{P}_{\tX}} \left[D\left( \mathbb{P}_{Y|X=\tX} \middle\| \mathbb{Q}_{Y|X=\tX}   \right)\right]$, 
where $(X,\tX)$ are independent and identically distributed (i.i.d.).
The mutual information between random variables $X$ and $Y$ is defined as $I(X;Y) \triangleq D\left( \mathbb{P}_{XY} \middle\| \mathbb{P}_X \mathbb{P}_Y \right)$.
The conditional mutual information between random variables $X$ and $Y$ given random variable $Z$ is defined as $I(X;Y|Z) \triangleq \mathbb{E}_{\mathbb{P}_{\tilde{Z}}}\left[I(X;Y|Z=\tilde{Z}) \right]$, 
where $(Z,\tZ)$ are i.i.d.

\section{Problem Setup}
\label{s:model}

We consider the networked control system depicted in \figref{fig:sns}, where the plant dynamics are described by a scalar, discrete-time, linear time-invariant (LTI) system
	\begin{align}
 	\label{eq:plant}
   		X_{k+1}=aX_k+U_k+W_k,
 	\end{align}
 	 where $X_k$, $a$, $U_k$, $W_k$ are real numbers representing the plant state, open-loop gain of the plant, control input, and plant disturbance, respectively, at time $k \in \nats$. The controller, at time $k$, observes $Y_k$
and generates a control signal $U_k$ as a function of $Y_1^k$. If the attacker does not tamper sensor reading, at any time $k \in \nats$, we have $Y_k=X_k$.  
We assume that the initial condition $X_0$ has a known (to all parties) distribution
   and is independent of the disturbance sequence $\{W_k\}$. For analytical purposes, we assume that the process $\{W_k\}$ has i.i.d.\ Gaussian samples of zero mean and variance $\sigma^2$  
   that is known to all parties. We assume, without loss of generality, that $W_0 = 0$, $\E{X_0}=0$,
   and take $U_0 = 0$. 
   Moreover, to simplify the notation, let $Z_k \triangleq (X_k, U_k)$ denote the state-and-control input at time $k$ and its trajectory up to time $k$---by 
\begin{align*}
	Z_1^k \triangleq (X_1^k, U_1^k) .
\end{align*}
The controller is equipped with a detector that tests for anomalies in the observed history $Y_1^k$.  When the controller detects an attack, it shuts the system down 
and prevents the attacker from causing further ``damage'' to the plant. The controller/detector is aware of the plant dynamics~\eqref{eq:plant} and knows  the open-loop gain $a$ of the plant.
On the other hand, the attacker knows the plant dynamics~\eqref{eq:plant} as well as the plant state $X_k$, and control input $U_k$ (or equivalently, $Z_k$) at time $k$ (see \figref{fig:sns}). However, it does not know the open-loop gain $a$\second{ of the plant}{}.

 We assume the open-loop gain is fixed in time, but unknown to the attacker (as in \blue{the} frequentist approach~\cite{efron_hastie_2016}).  Nevertheless,
it will be convenient, for algorithm aid, to assume a prior over the open-loop gain of the plant, and treat it as a random variable $A$, that is \textit{fixed across time}, whose PDF $f_A$ is known to the attacker, and whose realization $a$ is known to the controller (cf. \secref{ss:learning_controller}).
We assume all random variables 
to exist on a common probability space with probability measure $\mathbb{P}$, and $U_k$ to be  a \textit{measurable function} of $Y_1^k$ for all time $k \in \mathbb{N}$. We also denote the probability measure conditioned on $A = a$ by $\mathbb{P}_a$.
Namely, for any measurable event $C$, we~define
\begin{align*}
    \mathbb{P}_a(C) = \mathbb{P}(C|A=a).
\end{align*}
$A$ is 
assumed to be independent of $X_0$ and $\{W_k | k \in \nats\}$.


\subsection{Learning-based Attacks}
\label{ss:model:integrity_attack}

We now define \textit{Learning-based  attacks} that consist of two disjoint, consecutive, passive and    active phases (cf. \secref{ss:expbrffexploih}).

\textit{Phase 1: Learning.} 
During this phase, the attacker passively observes the control input and the plant state to learn
the open-loop gain of the plant. 
As illustrated in \figref{fig:sn1},
for all $k \in [0, L]$, 
the attacker observes the control input $U_{k}$ and the plant state $X_{k}$, 
and tries to learn the open-loop gain $a$, where $L$ is the duration of the learning phase. We denote by $\hA$ the attacker's estimate of the open-loop gain $a$.
\oprocend 

\textit{Phase 2: Hijacking.}
In this phase, the attacker aims to destroy the plant 
via 
$\tilde{U}_k$ while remaining undetected.
As~illustrated in \figref{fig:sn2}, 
from time $L+1$ and onward
the attacker~hijacks the system and feeds a malicious control signal $\tilde{U}_k$ to the plant and a fictitious sensor reading $Y_k = V_k$ to the controller.~\oprocend

We assume that the attacker can use \textit{any arbitrary} learning algorithm to estimate the open-loop gain $a$ during the learning phase, and \blue{when the} estimation is completed, we  assume that during the hijacking phase
the fictitious sensor reading
is constructed, in a model-based manner (cf. \secref{ss:basedvsfree}), as~follows
\begin{align}
\label{mimicmodel}
	V_{k+1} &= \hA V_k + U_k + \tW_k \,, & k = L, \ldots, T-1 ,
\end{align}
where
$\tW_k$ for $k = L, \ldots, T - 1$
are i.i.d.\ Gaussian $\mathcal{N}(0,\sigma^{2})$; 
$U_k$ is the control signal generated by the controller, 
which is fed with the fictitious virtual signal $V_k$ by the attacker; $V_L=X_L$; 
and $\hA$ is the estimate of the open-loop gain of the plant at the conclusion of Phase 1. 

\vspace{-0.75\baselineskip}
\subsection{Detection}
\label{ss:model:variance-test}

The controller/detector, being aware of the dynamic~\eqref{eq:plant} and the open-loop gain $a$, 
attempts to detect possible attacks by testing for statistical deviations from the typical behavior of the system~\eqref{eq:plant}.
More precisely, under the legitimate system operation (corresponding to the \textit{null hypothesis}), the controller observation $Y_k$ behaves according to 
\begin{align}
\label{L2Lshouldsatf}
	Y_{k+1} - a Y_k - U_k(Y_1^k) \sim ~\iid~ \mathcal{N}(0,\sigma^{2}).
\end{align}
In the case of an attack, during Phase 2 ($k > L$), 
\eqref{L2Lshouldsatf} can be rewritten as 
\begin{subequations}
\label{eq:subtract4test}
\noeqref{eq:substract4test:extended}
\begin{align}
 	V_{k+1}-aV_k-U_k
    &= V_{k+1} - aV_k + \hA V_k - \hA V_k-U_k \ \ 
\label{eq:substract4test:extended}
\\ &
    = \tW_k + \left( \hA - a \right) V_k,
\label{eq:substract4test:reduced}
\end{align}
\end{subequations}
where 
\eqref{eq:substract4test:reduced} 
follows from \eqref{mimicmodel}.
Hence, the estimation error $(\hA - a)$ dictates the ease 
with which an attack can be detected. 

Since 
the Gaussian PDF with zero mean is fully characterized by its variance, 
we shall follow~\cite{Kumar:DynamicWatermarki}, 
and test for anomalies in the latter,
i.e., test 
whether the empirical variance of \eqref{L2Lshouldsatf} is equal to the second moment of the plant disturbance $\E{W^2}$. To that end, we shall use a test that sets a confidence interval of length $2 \delta > 0$ around the expected variance, i.e., it checks whether
\begin{align}
\label{Test1}
\begin{aligned}
	&\frac{1}{T} \sum_{k=1}^{T} \left[ Y_{k+1}-a Y_k-U_k(Y_{1}^k) \right]^2 
 \\ &\qquad\qquad\qquad 
    \in (\Var{W}-\delta, \Var{W}+\delta),
\end{aligned}
\end{align}
where $T$ is called the \textit{test time}. That is, as is implied by \eqref{eq:subtract4test},
the attacker  deceives the controller and remains undetected 
if 
\begin{align*}
\begin{aligned}
\frac{1}{T} \left(\sum_{k=1}^{L} W_k^2+ \sum_{k=L+1}^{T} (\tW_k+(\hA-a)V_k)^2\right)
 \\  \in (\Var{W}-\delta, \Var{W}+\delta).
\end{aligned}
\end{align*}
\subsection{Performance Measures}
 \label{perf2039943!}
 
\begin{defn}
The hijack indicator at test time $T$ is defined as
\begin{align*}
	\Theta_T \triangleq
    \begin{cases}
        0, & \forall j \le T:\ Y_j = X_j \,;
	 \\ 1, & \textrm{otherwise} . 
    \end{cases}
\end{align*}
\blue{$\Theta_T$ is an oracle, and at} the test time $T$  the controller uses $Y_1^T$ to construct an estimate $\htheta_T$ of $\Theta_T$. More precisely, $\htheta_T=0$ if~\eqref{Test1} occurs, otherwise $\htheta_T=1$.\oprocend 
\end{defn}
\begin{defn}
The probability of deception is the  probability of the attacker deceiving the controller and remaining undetected at the time instant $T$
   \begin{align}
    \label{eq:dece_prob}
    	\Pdec^{a,T} \triangleq 				\mathbb{P}_a
        \left( \htheta_T=0 \middle| \Theta_T=1 \right) ;
    \end{align}
    the detection probability at test time $T$ is defined as 
    \begin{align*}
    	\Pdet^{a,T} \triangleq 
       1-\Pdec^{a,T}.
    \end{align*}
    Likewise, the probability of false alarm is the probability of detecting the attacker when it is not present, namely
    \begin{flalign*}
    	&&\qquad\qquad
    	\Pfa^{a,T} \triangleq 
        \mathbb{P}_a
        \left( \htheta_T=1 \middle| \Theta_T=0 \right). &&&& \ \ \oprocend
    \end{flalign*}
\end{defn}
Applying Chebyshev's inequality to \eqref{Test1} and noting that the system disturbances are i.i.d.\ Gaussian of variance $\sigma^2$,
we have 

\begin{align*}
\\[-2\baselineskip]
    \Pfa^{T} \le \frac{\mathrm{Var}[W^2]}{ \delta^2T}=\frac{3\sigma^4}{\delta^2T}.
\end{align*}
Further define the deception, detection, and false-alarm probabilities w.r.t.\ the probability measure $\mathbb{P}$, without conditioning on $A$, and denote them by $\Pdec^{T}$, $\Pdet^{T}$, and $\Pfa^{T}$, respectively. For instance, $\Pdet^T$ is defined, w.r.t.\ a PDF $f_A$ of $A$,~as 
\begin{align}
\label{eq:det-prob:explicit}
    \Pdet^{T} \triangleq \mathbb{P}\left( \htheta_T=1 \middle| \Theta_T=1 \right)
= \int_{-\infty}^{\infty} \Pdet^{a,T} \, f_A(a) da .
\end{align}


\vspace{-\baselineskip}
\section{Statement of the results}
\label{Statementwff2} 

In this section, we describe our main results  for the case of scalar plants. We provide lower and upper bounds on the deception probability \eqref{eq:dece_prob} of the learning-based attack \eqref{mimicmodel}, where the estimate $\hA$ in \eqref{mimicmodel} may be constructed using an \textit{arbitrary} learning algorithm. 
Our results are valid for \textit{any measurable} control policy $U_k$. 
We find a lower bound on the deception probability by characterizing what the attacker can at least achieve using a least-squares (LS) algorithm, and we derive  an information-theoretic converse for any learning algorithm using Fano's inequality~\cite[Chs.~2.10 \& 7.9]{cover2012elements}.  
While our analysis is restricted to the asymptotic case, $T \rightarrow \infty$, it is straightforward to extend 
it
to the non-asymptotic~case. 

For analytical purposes,
we assume that the power of the fictitious sensor reading is equal to $\beta^{-1} < \infty$, namely   
\begin{align}
\label{conversecss}
	\lim_{T\rightarrow \infty}&
	\frac{1}{T} \sum_{k=L+1}^{T} V_k^2
   	= 1 /\beta
    &\mbox{a.s. w.r.t.} ~\mathbb{P}_a.
\end{align}
\begin{remark}
\label{remark:aboutbeta2}
{\rm 
   Assuming the control policy is memoryless, namely $U_k$ is only dependent on $Y_k$, the process $V_k$ is Markov for $ k \ge L+1$. 
   By further assuming that $L=o(T)$ and using the generalization of the law of large numbers for Markov processes~\cite{durrett2010probability}, we deduce
   
   \begin{align*}
   \\[-2\baselineskip]
	\lim_{T\rightarrow \infty}
	&\frac{1}{T} \sum_{k=L+1}^{T} V_k^2
   	\ge \Var{W}
    &\mbox{a.s. w.r.t.} ~\mathbb{P}_a. 
\end{align*}
  Hence, in this case, we have $\beta \le 1/\Var{W}$. 
  Also, when the control policy is linear and stabilizes~\eqref{mimicmodel}, that is $U_k=-\Omega Y_k$ and $|\hA-\Omega|<1$, 
  it is easy to
verify that \eqref{conversecss} holds true for $\beta=(1-(\hA-\Omega)^2)/\Var{W}$. \blue{The assumption in \eqref{conversecss} can also be relaxed as described in Remarks~\ref{remark:afterrev3} and~\ref{remark:afterrev31}, in the sequel.}}~~\oprocend
\end{remark}

\blue{In the following lemma we show that for any learning-based attack~\eqref{mimicmodel}, as $T \to \infty$  the empirical variance used in the variance test~\eqref{Test1}  can be expressed in terms of the estimation error. The   result  follows from the strong law of large numbers applied to martingale difference sequences \cite[Lem.~2, part iii]{lai1982least}; it is   proved in \appref{lemprofr111113}.
\begin{lemma}
\label{Ineedtowk}
    \blue{Consider any} learning-based attack~\eqref{mimicmodel} and any measurable control policy $\{U_k\}$ such that the fictitious sensor reading power satisfies~\eqref{conversecss}.
    Then, the variance test~\eqref{Test1} reduces \blue{a.s., w.r.t.~$\mathbb{P}_a$,} to
    \begin{align}
        \lim_{T \rightarrow \infty} \frac{1}{T} \sum_{k=1}^{T} [Y_{k+1}-a Y_k-U_k(Y_{1}^k)]^2=
        \Var{W}+\frac{(\hA\!-\!a)^2}{\beta}.
    \nonumber
    \end{align}
\end{lemma}
}

\subsection{Lower Bound on the Deception Probability}
\label{sec:successful}

To provide a lower bound on the deception probability $\Pdec^{a,T}$, 
we consider a specific estimate $\hA$ at the conclusion of the first phase by the attacker.
%
Namely, we use LS estimation 
due to its efficiency and amenability to recursive update over observed incremental data~\cite{rantzer2018concentration,tu2017leastrecht,sarkar2019near}.
The LS algorithm approximates
the overdetermined system of equations
\begin{align*}
\begin{pmatrix}
  X_2 \\
  X_3 \\
  \vdots \\
  X_{L} \\
\end{pmatrix} =A \begin{pmatrix}
  X_1 \\
  X_2 \\
  \vdots \\
  X_{L-1} \\
\end{pmatrix}+\begin{pmatrix}
  U_1 \\
  U_2 \\
  \vdots \\
  U_{L-1} \\
\end{pmatrix},
\end{align*}
by minimizing the Euclidean distance $\hA =\argminA_A$
${\norm{X_{k+1}-A X_k - U_k}}$
to estimate (or ``identify'') the plant,
the solution to which is 

\begin{align}
\nonumber
\\[-2\baselineskip]
\label{learningAlgorithm}
	\hA &= \frac{\sum_{k=1}^{L-1}(X_{k+1}-U_k)X_k}{\sum_{k=1}^{L-1} X_k^2} \, 	
    &\mbox{a.s. w.r.t.} ~\mathbb{P}_a.
\end{align}

\begin{remark}{\rm 
    \eqref{learningAlgorithm} is well-defined
    since $\mathbb{P}_a(X_k = 0)=0$,
    as we assumed $W_k$ are i.i.d.\ zero-mean Gaussian 
    for all $k \in \mathbb{N}$.
     \oprocend}
\end{remark}


\blue{
We now lower bound   the deception probability of an attacker that utilizes LS estimation~\eqref{learningAlgorithm} under the variance test and in the presence of  any measurable  control policy for which~\eqref{conversecss} holds.
The following theorem demonstrates the  \textit{existence} of a learning-based attack that satisfies this lower bound.
As other learning algorithms may lead to better estimates, this also serves as a lower bound on the attacker's deception probability in the general case.
}

\begin{theorem}
\label{LB:dec-prob}
    \blue{Consider LS~\eqref{learningAlgorithm} learning-based attack~\eqref{mimicmodel} 
    and \blue{any measurable control policy $\{U_k\}$ such that the fictitious sensor readings satisfy~\eqref{conversecss}}.}
    Then,
    the 
    asymptotic deception probability under the variance test~\eqref{Test1} 
    is lower bounded as 
     \begin{subequations}
    \label{k2e1jmfve!!!!}
    \noeqref{k2e1jmfve!!!!1}
    \begin{align}
		    \lim_{T \rightarrow \infty} \Pdec^{a,T} 
            &=\mathbb{P}_a\left(|\hA-a|<\sqrt{\delta\beta}\right) 
            \label{k2e1jmfve!!!!1}
            \\
            \label{k2e1jmfve!!!!2}
            &\ge \mathbb{P}_a  \left( \frac{\left|\sum_{k=1}^{L-1}W_kX_k\right|}{\sum_{k=1}^{L-1} X_k^2} <\sqrt{\delta\beta} \right)
            \\
           &\ge 1-\frac{2}{(1+\delta\beta)^{L/2}}.
           \label{k2e1jmfve!!!!3}
    \end{align}
     \end{subequations}
\end{theorem}
\blue{
\begin{proof}
    \eqref{k2e1jmfve!!!!1} follows from \lemref{Ineedtowk}, and the dominated convergence theorem~\cite{durrett2010probability}. For  details see \appref{lemprofr11111344}. 
    Clearly, the estimation error of the LS algorithm~\eqref{learningAlgorithm} is \cite{rantzer2018concentration}
\begin{align}
\label{estsimerr222336227}
	\hA-a &= \frac{\sum_{k=1}^{L-1}W_kX_k}{\sum_{k=1}^{L-1} X_k^2} \, 	
     &\mbox{a.s. w.r.t.}~\mathbb{P}_a.
\end{align}
Consequently, by~\eqref{k2e1jmfve!!!!1}, a learning-based attack~\eqref{mimicmodel} can at least achieve the asymptotic deception probability~\eqref{k2e1jmfve!!!!2}. Finally,
    \eqref{k2e1jmfve!!!!3} holds by the concentration of measure   \cite[Th.~4]{rantzer2018concentration} by noting that $U_k$ is a measurable function of $Y_1^k = X_1^k$, for $k \in \{1,\ldots,L\}$.
%
\end{proof}
}

\begin{remark}
\label{remark:1linearcase333}
    {\rm 
    We can study the special case of Theorem~\ref{LB:dec-prob} for a linear controller. 
    Using the value of $\beta$ calculated in Remark~\ref{remark:aboutbeta2} for a linear controller $U_k=-\Omega Y_k$ when $|\hA-\Omega|<1$, we can rewrite~\eqref{k2e1jmfve!!!!3} as
    \begin{align}\label{LB:forlinearcon2}
        \lim_{T \rightarrow \infty} \Pdec^{a,T} \ge 1-\frac{2}{\left(1+\delta\frac{1-(\hA-\Omega)^2}{\Var{W}}\right)^{L/2}}.
    \end{align}
    In this case, for a fixed $L$, as $\Var{W}$ increases the lower bound in~\eqref{LB:forlinearcon2}  decreases. The reduction of the attacker's success rate can be explained by noticing that LS estimation~\eqref{learningAlgorithm} is based on minimizing $\norm{X_{k+1}-A X_k - U_k}$, and the precision of this estimate decreases as $\Var{W}$ increases.~\oprocend}
\end{remark}

\blue{\begin{remark}
\label{remark:afterrev3}
    {\rm 
    By replacing the limit with limsup in the   lower bound in~\thmref{LB:dec-prob}, the  result holds even if the limit in~\eqref{conversecss} does not exist. Also, if the limit in~\eqref{conversecss} is infinite then either the attacker or the
controller are doing a poor job, as described next. Assume the attacker uses an  estimate $\hat{A}$ such that at the conclusion of the learning phase  $\hat{A}$ belongs to the interval $(A-\delta',A+\delta')$, for a small value of $\delta'>0$. In this case, if the power of the fictitious sensor reading tends to infinity, then the controller is not robust. On the other hand, assume that the controller stabilizes the system with any open-loop gain that belongs to the interval $(A-\delta',A+\delta')$, where $\delta'>0$. In this case, if the power of fictitious sensor tends to infinity, then the attacker's estimate at the conclusion of the learning phase, does not belong to the interval $(A-\delta',A+\delta')$, i.e. the absolute value of the attacker's estimation error is larger than $\delta'$.}
    \end{remark}}

\blue{
\begin{exmp}
\label{ex:scalar:deception-rate}
      {\rm 
    In this example, we compare the empirical performance of the variance-test with our developed bound in~\thmref{LB:dec-prob}. At every time $T$,
    the controller
    tests the empirical variance for anomalies  over a detection window  $[1, T]$,
    using a confidence interval $2 \delta > 0$ around the expected variance~\eqref{Test1}. 
    %
    Here, $a=1$,  $\delta=0.1$, $U_k = -0.88aY_k$ for all $1\le k\le T=800$, and $\{W_k\}$ are i.i.d. Gaussian $\mathcal{N}(0,1)$, and 500 Monte Carlo simulations were performed. 
    
    The learning-based attacker~\eqref{mimicmodel} uses the LS algorithm~\eqref{learningAlgorithm} to estimate $a$  and, as illustrated in~\figref{fig:mnahayii!!!!!2213}, 
\begin{figure}[t]
\centering      
    \includegraphics[width = \columnwidth, trim = {0 0 0 2\baselineskip},clip]
    {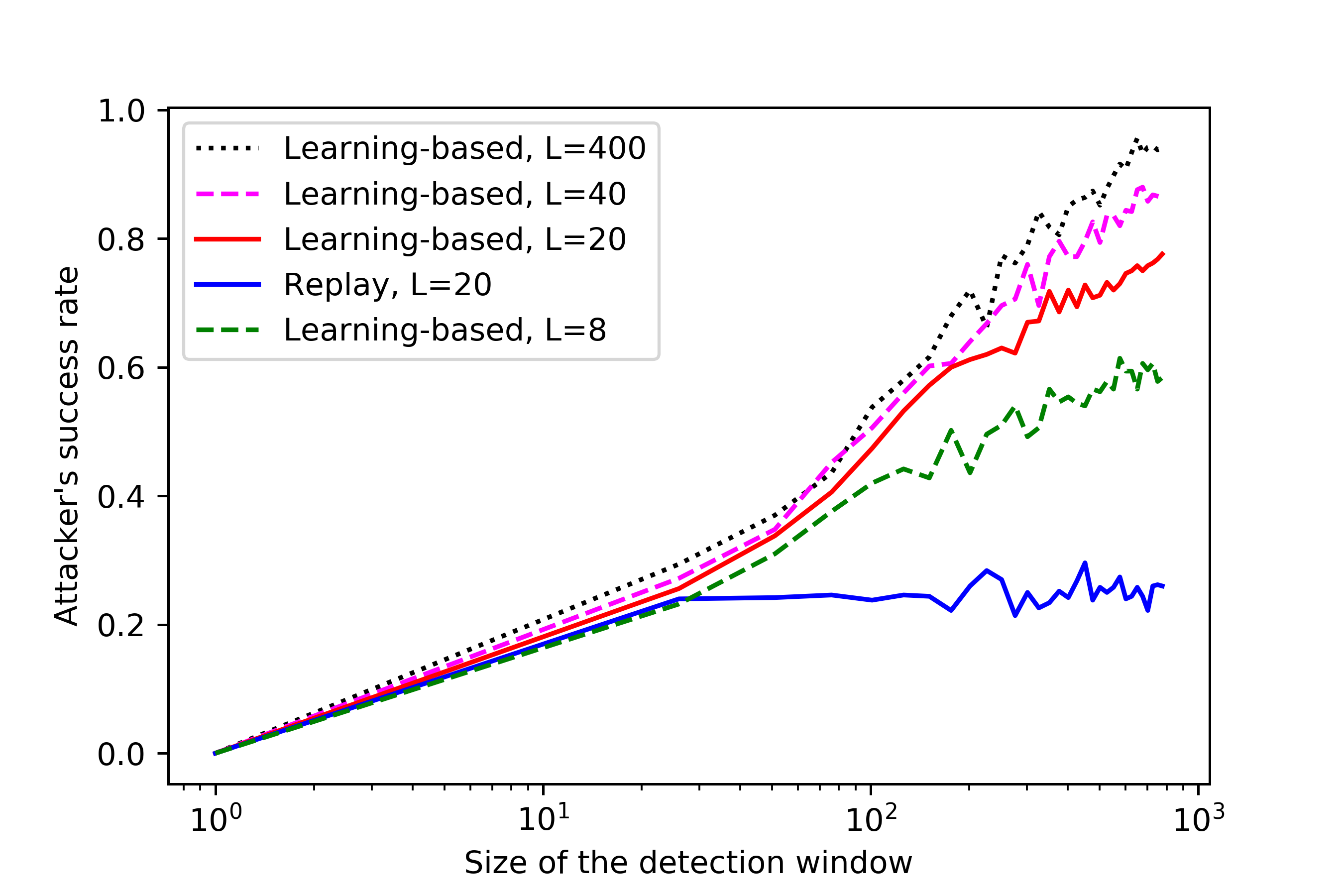}
        \caption{The attacker's success rate $\Pdec^{a,T}$ versus the size of the detection window $T$. 
    }
  \label{fig:mnahayii!!!!!2213}
\end{figure}
    the attacker's success rate increases as the duration of the learning phase $L$ increases. This is in agreement with~\eqref{k2e1jmfve!!!!3} since the attacker can improve its estimate of $a$ and the estimation error $|\hat{A}-a|$ reduces as $L$ increases. As discussed in~\secref{perf2039943!}, the false alarm rate decays to zero as the size of the detection window $T$ tends to infinity. 
    Hence, for a sufficiently large detection window, the attacker's success rate could potentially tend to one. Indeed, such behavior is observed in \figref{fig:mnahayii!!!!!2213} for a learning-based attacker~\eqref{mimicmodel} with $L=400$.
    \figref{fig:mnahayii!!!!!2213} also illustrates that our learning-based attack  outperforms  the replay attack. A replay attack with a recording length of $L=20$ and a learning-based attack with a learning phase of length $L=20$ are compared, and the success rate of the replay attack saturates at a lower value. 
    Moreover, a learning-based attack with a learning phase of length $L=8$ has a higher success rate than a replay attack with a larger recording length of $L=20$.}
    \oprocend
\end{exmp}
}


\subsection{Upper Bound on the Deception Probability}
\label{mmlfle2}
We now derive an upper bound on the deception probability
~\eqref{eq:dece_prob} of any learning-based attack~\eqref{mimicmodel} where $\hA$ in \eqref{mimicmodel} is constructed using \textit{any arbitrary} learning algorithm, for \textit{any measurable} control policy,
when  $A$ is distributed over a symmetric interval $[-R, R]$.   
Since the uniform distribution has the highest entropy 
among all distributions with finite support~\cite[Ch.~12]{cover2012elements}, we assume that $A$ has a uniform prior over the interval $[-R, R]$.
We further assume that the attacker knows this distribution (including the value of $R$), whereas the controller knows the true realization of $A$ (as before). 
\begin{theorem}
\label{conversew}
	For any $R > 0$, let $A$ be distributed uniformly over $[-R, R]$, and consider \textit{any measurable} control policy $\{U_k\}$ and \textit{any learning-based attack}~\eqref{mimicmodel} 
    \blue{such that} the fictitious sensor readings satisfy~\eqref{conversecss} with $\sqrt{\delta\beta} \le R$.
	Then, the asymptotic deception probability, when using the variance test~\eqref{Test1}, is upper bounded   as 
%
    \begin{subequations}
    \label{k2e1jmfve}
    \noeqref{k2e1jmfve:lim,k2e1jmfve:UB}
    \begin{align}
        \lim_{T \rightarrow \infty}\Pdec^{T} 
	 &=\mathbb{P}(|A-\hA|< \sqrt{\delta\beta})
    \label{k2e1jmfve:lim}
    \\ &\le 
      \Lambda\triangleq \dfrac{I(A;Z_1^L)+1}{\log (R/\sqrt{\delta\beta})} \,. 
    \label{k2e1jmfve:UB}
    \end{align}
    \end{subequations}
In addition, if for all $k \in \{1,\ldots,L\}$, $A\rightarrow  (X_k,Z_1^{k-1}) \rightarrow U_k$ is a Markov chain, then for any  sequence of probability measures $\{ \mathbb{Q}_{X_k|Z_1^{k-1}} \}$,
such that for all $k \in \{ 1, \ldots, L \}$ $\mathbb{P}_{X_k|Z_1^{k-1}}\ll \mathbb{Q}_{X_k|Z_1^{k-1}}$, we have 
    \begin{align} 
    \label{newlowreddw}
      &\Lambda
    \le
       \dfrac{\sum_{k=1}^{L} D\left(\mathbb{P}_{X_k|Z_1^{k-1},A} \middle\| \mathbb{Q}_{X_k|Z_1^{k-1}} \middle| \mathbb{P}_{Z_1^{k-1},A} \right) + 1}{\log \left( R/\sqrt[]{\delta\beta} \right)}\,. \ \ 
    \end{align}

%
\end{theorem}
\blue{
\begin{proof}
    \eqref{k2e1jmfve:lim} follows from \eqref{eq:det-prob:explicit},~\eqref{k2e1jmfve!!!!1}, and Tonelli's theorem ~\cite{durrett2010probability}.
    For   details see~\appref{proof13as}.
%
\eqref{k2e1jmfve:UB} follows by noting that 
since the attacker observes the plant state and control input during the learning phase which lasts $L$ steps, and since $A\rightarrow  (X_1^L,U_1^L) \rightarrow \hA$ constitutes a Markov chain, using the continuous domain version of Fano's inequality~\cite[Prop.~2]{duchi2013distance}, we have
\begin{align}
\label{Fanoforcontin}
    \inf_{\hA} \mathbb{P} \left( \left| \hA - A \right| \ge \sqrt{\delta\beta} \right) 
    \ge 1-\frac{I(A;Z_1^L)+1}{\log (R/\sqrt{\delta\beta})},
\end{align}
whenever ${\sqrt{\delta\beta} \le R}$. 
%
Finally, \eqref{newlowreddw} follows the arguments of \cite{raginsky2010divergence}
and is proven, for completeness, in \appref{app:KL:manipulations}.
\end{proof}
}
\begin{remark}
\label{rmark12} 
{\rm 
By looking at the numerator in \eqref{k2e1jmfve:UB}, it follows that the bound on the deception probability becomes looser as  the amount of information revealed about the open-loop gain $A$ by the observation $Z_1^L$ increases. On the other hand, by looking at the denominator, the bound becomes tighter as $R$ increases. This is consistent with the observation of Zames~\cite{raginsky2010divergence} 
that system
identification becomes harder as the uncertainty about the open-loop gain of the plant increases. 
    In our case, a larger uncertainty interval $R$ corresponds to a poorer estimation of 
    $A$ by the attacker, which leads, in turn, 
    to a decrease in the 
    achievable deception probability. 
    The denominator can  also be interpreted as the intrinsic uncertainty of $A$ when it is observed at resolution~$\sqrt{\delta \beta}$, as it corresponds to the entropy of the random variable $A$ when it is quantized at such resolution.} \oprocend
\end{remark}

In conclusion, \thmref{conversew} provides two upper bounds on the deception probability. 
The first bound~\eqref{k2e1jmfve:UB} clearly shows that 
increasing the privacy of the open-loop gain $A$---manifested in the mutual information between $A$ and the state-and-control trajectory $Z_1^L$ during the exploration phase---reduces the deception probability.
The second bound~\eqref{newlowreddw} allows freedom in choosing the auxiliary probability measure $\mathbb{Q}_{X_k|Z_1^{k-1}}$, making it a rather useful bound.
For instance, by choosing 
$\mathbb{Q}_{X_k|Z_1^{k-1}}\sim \mathcal{N}(0,\sigma^{2})$, 
for all $k \in \nats$, we can rewrite the upper bound~\eqref{newlowreddw} in term of $\Epi{(AX_{k-1}+U_{k-1})^2}$ as follows. The proof of the following corollary can be found in the appendix.
\begin{corollary}
\label{uplowthem}
 	Under the assumptions of \thmref{conversew}, 
    if  for all $k \in \{1,\ldots,L\}$,  $A\rightarrow  (X_k,Z_1^{k-1}) \rightarrow U_k$ is a Markov chain, 
 	then the
 	asymptotic deception probability upper bounded as
 	\begin{subequations}
 	\begin{align} 
    \label{newlowreddxssasw}
                &\quad\quad\quad\quad\quad\quad
                \lim_{T \rightarrow \infty}\Pdec^{T}
                \le  G(Z_1^L),
\\
\label{gbarupper1}
   &G(Z_1^L) \triangleq        \dfrac{\frac{\log e}{2\sigma^2}\sum_{k=1}^L \Epi{(AX_{k-1}+U_{k-1})^2}+1}{\log \left( R / \sqrt[]{\delta\beta}\right)} \,.~\, 
   \end{align}
 	\end{subequations}
\end{corollary}
\blue{\begin{remark}
\label{remark:afterrev31}
    {\rm 
    Following  the same discussion as in Remark~\ref{remark:afterrev3}, by replacing the limit with liminf in the upper bound in~\thmref{conversew}, the derived results remain true even if~\eqref{conversecss} does not happen.}
    \oprocend
    \end{remark}}
\blue{The next example} compares the lower and upper bounds on the deception probability of \thmref{LB:dec-prob} and \colref{uplowthem}.

\blue{
\begin{exmp}
\label{ex:scalar:LB-UB-compare}
    {\rm 
    \thmref{LB:dec-prob} provides a lower bound on the deception probability 
    given
    $A=a$. 
    Hence, by
    applying the law of total probability
    w.r.t.\ the PDF $f_A$   as in~\eqref{eq:det-prob:explicit}, we can apply the result of \thmref{LB:dec-prob} to provide a lower bound also on the average deception probability for a random open-loop gain $A$. 
    In this context, \figref{fig:cdmkcdmk11} 
\begin{figure}[t]
\centering
	  \includegraphics[width = \columnwidth, trim = {0 0 0 2.6\baselineskip}, clip]
	  {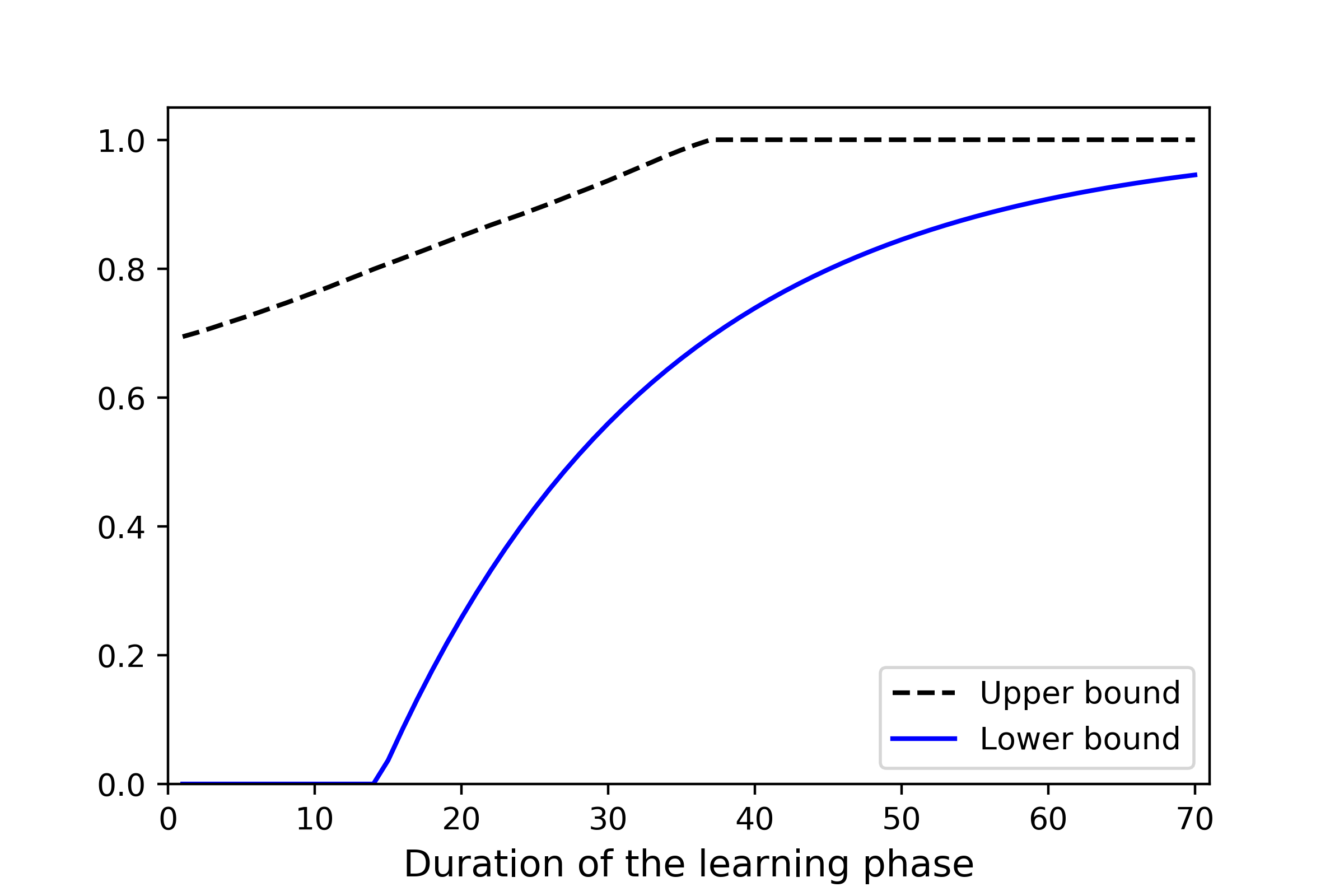}
        \caption{
        Comparison of the lower and upper bounds on the deception probability, of \thmref{LB:dec-prob} and \colref{uplowthem}, respectively.}
        \label{fig:cdmkcdmk11}
\end{figure}    
    compares the  lower and upper bounds on the deception probability provided by \thmref{LB:dec-prob} and  
    \colref{uplowthem}, augmented with the trivial cases of zero and one probability, namely $\max\{0,1-(2/(1+\delta\beta)^{L/2})\}$, and $\min\{ 1,G(Z_1^L)\}$,  where $A$ is distributed uniformly over $[-0.9, 0.9]$. Equation \eqref{newlowreddxssasw} is valid when the control input is not a function of random variable $A$; hence, we assumed $U_k = -0.045Y_k$ for all time $k \in \mathbb{N}$.
        Here $\delta=0.1$, $\{W_k\}$ are i.i.d. Gaussian with zero mean and variance of $0.16$, and for simplicity, we assume the limit in~\eqref{conversecss} exists [cf. Remarks~\ref{remark:afterrev3} and~\ref{remark:afterrev31}], and we let $\beta=1.1$. Although  in general  the attacker's estimation of the random open-loop gain $A$ and consequently the power of fictitious sensor reading~\eqref{conversecss} vary based on the learning algorithm and the realization of $A$, the comparison of  the  lower and upper bounds in~\figref{fig:cdmkcdmk11} is restricted to a fixed $\beta$.  
        $2000$ Monte Carlo simulations were performed.
        
        \figref{fig:cdmkcdmk11} also illustrates the gap between these  lower and upper bounds on the deception probability.
        By restricting the class of control policies or learning algorithms one might be able to derive tighter results at the cost of losing generality.
        }
        \oprocend
\end{exmp}
}
\subsection{Privacy-enhancing Signal}
\label{subsectionauthernt11} 

For a given duration of the learning phase $L$, to increase the security of the system, at any time $k$  
the controller can add a privacy-enhancing signal $\Gamma_k$ 
to an unauthenticated
control policy $\{\oU_k | k \in \nats\}$:
\begin{align}
\label{eq:watermaked_control}
	U_k &= \oU_k + \Gamma_k \,, & k \in \nats .
\end{align}
We refer to such a control policy~$U_k$ as the \textit{authenticated} 
control policy~$\oU_k$.
We denote the states of the system that would be generated if only the unauthenticated control signal~$\oU_1^k$ were applied by~$\oX_1^k$, 
and the resulting trajectory---by $\oZ_1^k \triangleq (\oX_1^k,\oU_1^k)$. 

\blue{The following numerical example illustrates the effect of the privacy-enhancing signal on the deception probability.
\begin{exmp}
\label{ex:authentication}
  {\rm 
  Here, the attacker uses the LS algorithm~\eqref{learningAlgorithm}, the detector uses the variance test~\eqref{Test1}, $a=1$, $T=600$, $\delta=0.1$, and $\{W_k\}$ are i.i.d. standard Gaussian. 
We
compare 
the attacker's success rate, the empirical $\Pdec^{a,T}$,
as a function of the duration $L$ of the learning phase
for three different control policies: 
I) unauthenticated control signal~$\oU_1^k=-aY_k$ for all $k$,  II) authenticated control signal~\eqref{eq:watermaked_control}, where $\Gamma_k$ are i.i.d.\ Gaussian $\mathcal{N}(0,9)$,  III) authenticated control signal~\eqref{eq:watermaked_control}, where $\Gamma_k$ are i.i.d.\ Gaussian $\mathcal{N}(0,16)$. As illustrated in~\figref{fig:mnahayiii2222}, 
\begin{figure}[t]
 \centering      
     \includegraphics[width=\columnwidth, trim = {0 0 0 2.6\baselineskip}, clip]
     {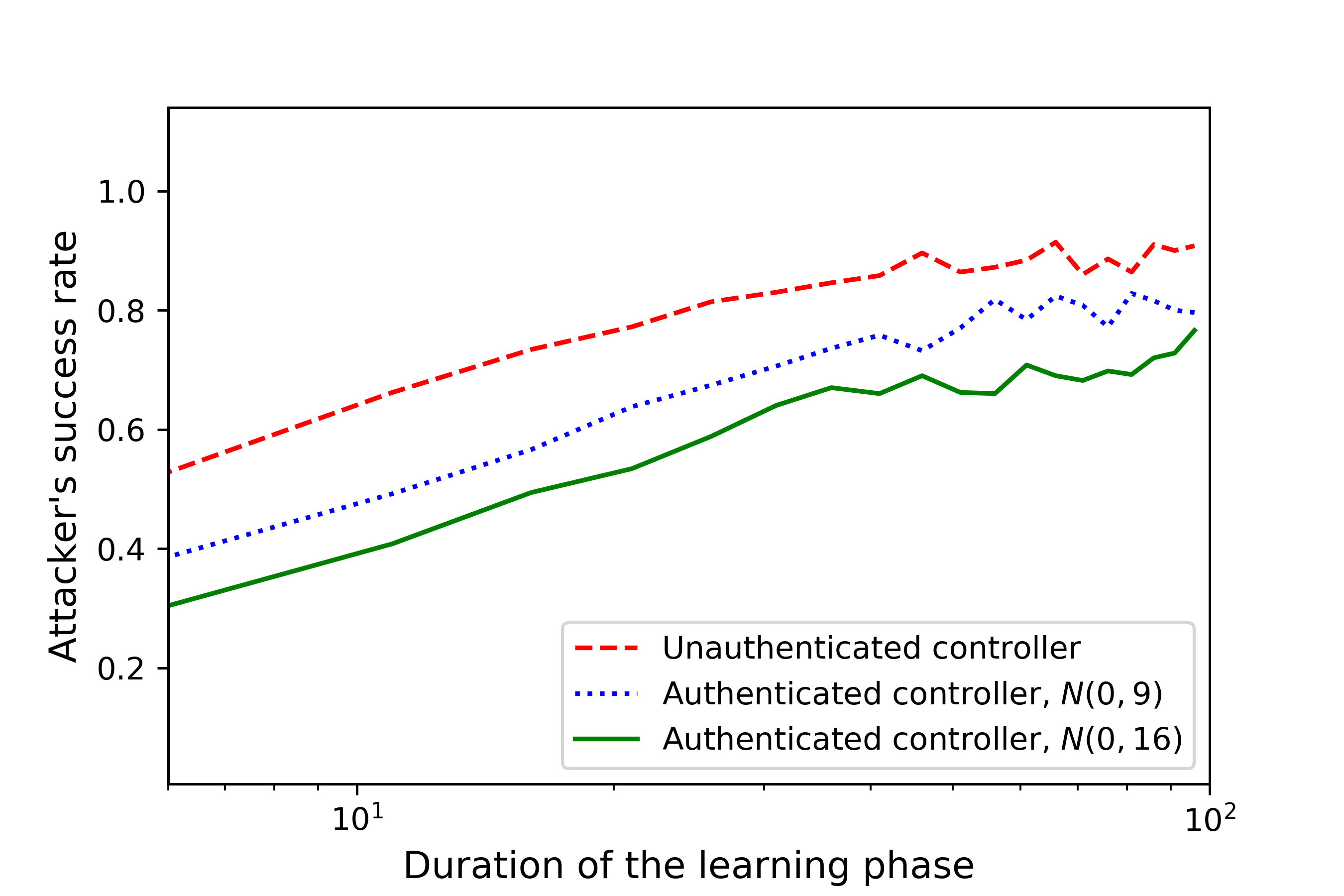}
         \caption{The attacker's success rate $\Pdec^{a,T}$ versus the duration of the learning phase $L$.}
\vspace{-\baselineskip}
   \label{fig:mnahayiii2222}
\end{figure}
for both the authenticated and unauthenticated control signals, the attacker's success rate increases as the duration of the learning phase increases. This is in agreement with~\eqref{k2e1jmfve!!!!3} since the attacker can improve its estimate of $a$ as $L$ increases. Further, for a fixed $L$
the attacker performance deteriorates as the power of privacy-enhancing signal $\Gamma_k$ increases. Namely, $\Gamma_k$ hampers the learning process of the attacker and the estimation error $|\hat{A}-a|$ increases as the power of the privacy-enhancing signal increases. 500 Monte Carlo simulations were performed.} 
\oprocend
\end{exmp}}

\begin{remark}
\label{rem405322!}
{\rm 
   A ``good'' privacy-enhancing signal entails little increase in the control cost~\cite{bertsekas2019reinforcement} compared to its unauthenticated version
while providing enhanced detection probability~\eqref{eq:dece_prob} and/or false alarm probability. Finding the optimal privacy-enhancing signal is left for future research. 
\blue{We remark that since the submission of this
paper~\cite{supplementaryus,khojasteh2019authentication}, some latter literature has appeared that builds
on it. In particular,}
in a follow-up work~\cite{follow-up}, Ziemann and Sandberg \blue{focused on
designing optimal privacy-enhancing signal,} by studying the  optimal control problem of linear systems regularized with Fisher
information, 
where the latter serves as a proxy to the estimation quality of $A$ via the Cramer--Rao lower bound. \oprocend}
\end{remark}

One may envisage that superimposing any noisy signal $\Gamma_k$ on top of the control policy $\{\oU_k | k \in \nats\}$ would necessarily enhance the detectability of \textit{any} learning-based attack~\eqref{mimicmodel}
since the  observations of the attacker are in this case noisier.
However, it turns out that injecting a strong noise for some learning algorithm may, in fact, speed up the learning process as it improves the power of the signal magnified by the open-loop gains with respect to the observed noise~\cite{rantzer2018concentration}.
Any signal $\Gamma_k$ that satisfies the condition proposed in the following corollary, whose proof available in the appendix,
will
provide enhanced guarantees on the detection probability when the attacker uses \textit{any arbitrary} learning algorithm to estimate the uniformly distributed $A$ over the symmetric interval $[-R, R]$. 
\begin{corollary}
\label{last2efthm1}
For any control policy $\{\oU_k | k \in \nats\}$ 
with trajectory $\oZ_1^k=(\oX_1^k,\oU_1^k)$ and its corresponding authenticated control policy $U_1^k$~\eqref{eq:watermaked_control} with trajectory $Z_1^k= (X_1^k,U_1^k)$,  under the assumptions of Corollary~\ref{uplowthem},  if for all 
$k \in \{1,\ldots,L-1\}$ 
  \begin{align}\label{expt431mms}
    	\Epi{\Psi_k^2+2\Psi_k(A\bar{X}_k+\bar{U}_k)} <0,
    \end{align} 
    where  
 $\Psi_k \triangleq \sum_{j=1}^k A^{k-j}\Gamma_j$,
 for any $L \ge 2$, 
    the following majorization of $G$~\eqref{gbarupper1} holds:
     \begin{align}
     \label{nkkie!!!3334}
    G(Z_1^L) < G(\bar{Z}_1^L).
    \end{align}
\end{corollary}

\blue{
\begin{remark}
    \colref{last2efthm1} can be generalized by replacing the limit with liminf in~\eqref{conversecss} [cf.  Remarks~\ref{remark:afterrev3} and~\ref{remark:afterrev31}].
    \oprocend
\end{remark}}

\begin{exmp}
\label{re:rr444exap13}
In this example, we describe a class of privacy-enhancing signals that yield better guarantees on the deception probability. For all $k \in \{2,\ldots,L\}$, clearly, 
$\Psi_{k-1}=-(A\oX_{k-1}+\oU_{k-1})/\eta$ satisfies the condition in~\eqref{expt431mms} for any $\eta \ge 3$. 
Thus, by choosing the privacy-enhancing signals  $\Gamma_{1}=-(A\oX_{1}+\oU_{1})/\eta$, and $\Gamma_{k}=-(A\oX_{k}+\oU_{k})/\eta-\sum_{j=1}^{k-1}A^{k-1-j}\Gamma_{j}$ 
for all $k \in \{3,\ldots,L\}$, 
~\eqref{nkkie!!!3334}~holds. \blue{A numerical example for this authentication policy  demonstrates a decrease in the deception probability at the expense of a higher control cost, and it can be found in~\appref{ExamplenumLQR44111}.
}
\oprocend
\end{exmp}
\blue{
\begin{remark}
\label{remark:watme444}
    {\rm The privacy-enhancing signal introduced in this work is  related to the dynamic watermarking signal~\cite{MoSinopoli:Magzine,hespanhol2018statistical,Kumar:DynamicWatermarki}, which are unique signatures that are  available \textit{only} to the controller.  In contrast, in our setup (depicted in~\figref{fig:sn2}), the attacker has access to the
signal generated by the controller. Thus, by reading the control input at time $k$ and  constructing the ficticous sensor reading $V_{k+1}$, as in~\eqref{mimicmodel}, the attacker can construct a fictitious sensor reading  containing any watermark signal inscribed by the controller. It follows that techniques based on dynamic watermarking that rely on the privacy of such signal break down in the case where attacks   have access to the control signal generated by the controller. 
Instead, we  take advantage of the authentication signal in a different way: since the attacker does not have full knowledge about the system dynamics, this signal is used to hamper the learning process of the attacker during the learning phase. }
\oprocend
    \end{remark}
    }

\section{Extension to Vector Systems}
\label{sec:vector2}

We now generalize   our results to vector systems.
Consider the networked control system depicted in \figref{fig:sns}, with the plant dynamics replaced by a \textit{vector} plant: 
	\begin{align}
 	\label{eq:plant-vec}
   		\textbf{x}_{k+1}=\texttt{A}\textbf{x}_k+\textbf{u}_k+\textbf{w}_k,
 	\end{align}
 	where $\textbf{x}_k \in \real^{n \times 1}$, $\textbf{u}_k \in \real^{n \times 1}$, $\texttt{A}  \in \real^{n \times n}$, $\bW_k \in \real^{n \times 1}$ represent the plant state, control input, open-loop gain of the plant, and plant disturbance, respectively, at time $k \in \nats$. The controller, at time $k$, observes $\textbf{y}_k$
and generates a control signal $\textbf{u}_k$ as a function of $\textbf{y}_1^k$, and $\textbf{y}_k=\textbf{x}_k$ at times $k \in \nats$ at which the attacker does not tamper the sensor reading.
 	We assume that the initial condition $\textbf{x}_0$ has a known (to all parties) distribution 
   and is independent of the disturbance sequence $\{\textbf{w}_k\}$. 
   For analytical purposes, we further assume $\{\textbf{w}_k\}$ is a process with i.i.d.\ multivariate Gaussian samples of zero mean and a covariance matrix $\bSigma$ that is known to all parties.
Without loss of generality, we assume that $\textbf{w}_0 = 0$, $\E{\textbf{x}_0}=0$, and take $\textbf{u}_0 = 0$.  
 	
 We assume the attacker uses the vector analogue of learning based attacks described in \secref{ss:model:integrity_attack} where
 the attacker can use \textit{any} learning algorithm to estimate the open-loop gain matrix $\texttt{A}$ during the learning phase.
 The estimation $\bhA$ constructed by the attacker at the conclusion of the learning phase is utilized to construct the fictitious sensor readings $\{\textbf{v}_k\}$ according to the vector analogue of~\eqref{mimicmodel}, where
$\{\btW_k | k = L, \ldots, T - 1\}$
are i.i.d.\ multivariate Gaussian with zero mean and covariance matrix $\bSigma$. 

Similar  to the scalar case, for analytical purposes,
we assume that the power of the fictitious sensor reading is equal to \mbox{$1/\beta < \infty$} [cf. Remarks~\ref{remark:aboutbeta2} and~\ref{remark:afterrev3}], namely   
\begin{align}
\label{conversecss_vec}
	\lim_{T\rightarrow \infty}&
	\frac{1}{T} \sum_{k=L+1}^{T} \|\bV_k\|^2
   	= \frac{1}{\beta}
    &\mbox{a.s. w.r.t.}~\mathbb{P}_\texttt{A} \,.
\end{align} 
 	
Since the zero-mean multivariate Gaussian distribution is completely characterized by its covariance matrix, we shall follow \cite{Kumar:DynamicWatermarki} and test for anomalies in the latter. 
To that end, define the error matrix
\begin{align}
	&\bDelta \triangleq \bSigma
    -\frac{1}{T} \sum_{k=1}^{T} \left[ \textbf{y}_{k+1}-\texttt{A} \textbf{y}_k-\textbf{u}_k \right]    \left[ \textbf{y}_{k+1}-\texttt{A} \textbf{y}_k-\textbf{u}_k  \right]^\dagger .
 \nonumber
\end{align}

As in~\eqref{Test1}, we use a test that sets a confidence interval, with respect to the norm, around the expected  covariance matrix, i.e., it checks whether
\begin{align}
\label{Testvec}
\|\bDelta\|_{op} \le \gamma,
\end{align}
at the test time $T$. 
For the sake of analysis, we use the operator norm in \eqref{Testvec}, which satisfies the sub-multiplicativity property.

The following lemma provides a necessary and sufficient condition for any learning-based attack [the vector analogue of \eqref{mimicmodel}] to deceive the controller and remain undetected, 
for a multivariate plant~\eqref{eq:plant-vec} under a covariance test~\eqref{Testvec}, 
in the limit of $T \to \infty$; its proof is available in the appendix.

\begin{lemma}
\label{lem:vecdetel}
    Consider the multivariate plant~\eqref{eq:plant-vec}, and any learning-based attack  analogous to \eqref{mimicmodel}, 
    with fictitious sensor reading power that satisfies~\eqref{conversecss_vec}, and any measurable control policy {\rm{$\{\textbf{u}_k\}$}}.
    Then, the attacker can deceive the controller and remain undetected, under the covariance test~\eqref{Testvec},  a.s.\ in the limit $T \rightarrow \infty$, if and only if 
    \rm{
        \begin{align}
     \label{feji-vec3}
         \lim_{T\rightarrow \infty}\frac{1}{T}\norm{\sum_{k=L+1}^{T}
    (\bhA-\texttt{A})\textbf{v}_k \textbf{v}_k^{\dagger}(\bhA-\texttt{A})^{\dagger}}_{op} \le \gamma .
     \end{align}
     }
\end{lemma}

\lemref{lem:vecdetel} has the following important implication.
\begin{subequations}
\label{implici-vec3}
\noeqref{implici-vec3:triangEQ,implici-vec3:norm,implici-vec3:power}
\begin{align}
         \lim_{T\rightarrow \infty}&\frac{1}{T}\norm{ \sum_{k=L+1}^{T}
    (\bhA-\texttt{A})\textbf{v}_k \textbf{v}_k^{\dagger}(\bhA-\texttt{A})^{\dagger}}_{op} 
    \label{implici-vec3:basic}
    \\
    &\le 
    \lim_{T\rightarrow \infty}\frac{1}{T} \sum_{k=L+1}^{T} \norm{
    (\bhA-\texttt{A})\textbf{v}_k ((\bhA-\texttt{A})\textbf{v}_k)^{\dagger}}_{op}
    \label{implici-vec3:triangEQ}\\
    &\le
    \lim_{T\rightarrow \infty}\frac{1}{T} \sum_{k=L+1}^{T}
    \norm{(\bhA-\texttt{A})\textbf{v}_k}_{op}^2 
    \label{implici-vec3:norm}\\
    & \le \norm{\bhA-\texttt{A}}_{op}^2 / \beta \,,
\label{implici-vec3:power}
\end{align}
\end{subequations}
 where \eqref{implici-vec3:triangEQ} follows from the triangle inequality, 
 \eqref{implici-vec3:norm} and \eqref{implici-vec3:power} follow from the sub-multiplicativity of the operator, the identity $\|\bV_k\|=\|\bV_k\|_{op}$ and by putting the power constraint \eqref{conversecss_vec} into force. 
 
 If $\|\bhA-\texttt{A}\|_{op}^2 \le \gamma\beta $, \eqref{feji-vec3} holds in the limit of $T \to \infty$, then the attacker is able to deceive the controller and remain undetected  a.s.,
 by  \lemref{lem:vecdetel}.
Equation \eqref{implici-vec3}  then  implies that the norm of the estimation error, $\|\bhA-\texttt{A}\|_{op}$, dictates the ease 
with which an attack can go undetected. 
This is used
next 
to develop a lower bound on the deception probability.

\subsection{Lower Bound on the Deception Probability}

We start by observing that in the case of multivariate systems, and in contrast to their scalar counterparts, some control actions might not reveal the entire plant dynamics, and in this case the attacker might not be able to learn the plant completely. 
This phenomenon is captured by the persistent excitation property of control inputs, 
which describes control-action signals that are 
sufficiently rich to excite all the system modes that will allow to learn them.
While \textit{avoiding} persistently exciting control inputs can be used as a way to \textit{secure} the system against learning-based attacks, here, we assume a probabilistic variant of this property~\cite{duflo2013random,raginsky2010divergence}. 

\begin{defn}[Persistent excitation]
\label{def:persistent_excitation}
    Given a plant~\eqref{eq:plant-vec}, $\zeta > 0$, and  $\rho \in [0,1]$, the control policy $\textbf{u}_k$ is $(\zeta,\rho)$-persistently exciting if there exists a time $L_0 \in \nats$  such that,  for all $\tau \ge L_0$, 
    \begin{align}\label{prexc345}
        \mathbb{P}_{\texttt{A}} \left(\frac{1}{\tau} \bG_{\tau}\succeq \zeta \Id_{n \times n}  \right)\ge \rho, 
    \end{align}
    where $\bG_\tau$ is the sum of the state Gramians up to time $\tau$:
    \begin{align}
       \label{Gram_matrix_1} \bG_\tau\triangleq\sum_{k=1}^{\tau} \bX_k \bX_k^{\dagger}.
    \end{align}
\end{defn}

As in \secref{sec:successful}, to find a lower bound on the deception probability $\Pdec^{\texttt{A},T}$, 
we consider a specific estimate of $\texttt{A}$, obtained via the LS estimation algorithm, analogous to \eqref{learningAlgorithm}, at the conclusion of the first phase by the attacker:
%
%
\begin{align}
\label{learningAlgorithmvect-var}
	\bhA =     
	\begin{cases}
        \bzero_{n \times n}, & \mbox{det}(\bG_{L-1})=0;
	 \\ 
	 \sum\limits_{k=1}^{L-1}(\bX_{k+1}-\bU_k)\bX^{\dagger}_k \bG_{L-1}^{-1}, & \textrm{otherwise} ,
    \end{cases} 
    \:
\end{align}
where $\bzero_{k \times \ell}$ denotes an all zero matrix of dimensions $k \times \ell$.

Next, we show an upper bound on the estimation-error norm, $\|\bhA-\texttt{A}\|_{op}$, of the above LS algorithm~\eqref{learningAlgorithmvect-var}, and use it to extend the bound in~\eqref{k2e1jmfve!!!!} to the vector case. A complete proof is available in the appendix.


\begin{lemma}
\label{lem:Raginskyerror}
   Consider the vector plant~\eqref{eq:plant-vec}. 
   If
   the attacker constructs {\rm{$\bhA$}} using LS estimation \eqref{learningAlgorithmvect-var}, and the controller uses a policy $\{\bU_k\}$ for which the event in~\eqref{prexc345} occurs for $\tau = L-1$, that is $\bG_{L-1} /(L-1) \succeq \zeta \texttt{I}_{n \times n}$.  
    Then, we have
    \rm{
    \begin{align}\label{upperlearning223-first}
    \|\bhA-\texttt{A}\|_{op}
    &\le \frac{1}{\zeta L}\sum_{k=1}^{L-1}\|\textbf{w}_k\textbf{x}_k^{\dagger}\|_{op}
    &\mbox{a.s. w.r.t.}~\mathbb{P}_\texttt{A} \,.
    \end{align}
    }
\end{lemma}
The following theorem provides a lower bound on the deception probability of an attacker that utilizes LS estimation~\eqref{learningAlgorithmvect-var}, and its proof can be found in the appendix. \blue{As discussed before \thmref{LB:dec-prob}, since the attacker might be able to construct better estimates using other learning algorithms, this also serves as a lower bound on the attacker's deception probability in the general case.}
\begin{theorem}
\label{LB:dec-prob-vec}
Consider the plant~\eqref{eq:plant-vec} with a $(\zeta,\rho)$-persistently exciting control policy {\rm{$\{\textbf{U}_k\}$}} from time $L_0$, and LS~\eqref{learningAlgorithmvect-var} learning-based attack [the vector analogue of~\eqref{mimicmodel}] \blue{such that} the fictitious sensor reading power satisfies~\eqref{conversecss_vec} and with a learning phase of duration $L \ge L_0+1$.
    Then, the asymptotic deception probability, when using the covariance test~\eqref{Testvec},
    is bounded from below as 
    \rm{
     \begin{subequations}
    \label{k!!vec!}
    \begin{align}
		    \lim_{T \rightarrow \infty} \Pdec^{\texttt{A},T} 
            &\ge \mathbb{P}_\texttt{A}\left(\|\bhA-\texttt{A}\|_{op}<\sqrt[]{\gamma\beta}\right)
            \label{k2e1jmfve!!!!1vec}
            \\
            &\ge \rho \mathbb{P}_\texttt{A}  \left(\frac{1}{\zeta L} \sum_{k=1}^{L-1}\|\textbf{w}_k\textbf{x}_k^{\dagger}\|_{op}<  \sqrt{\gamma\beta} \right) . \quad
    \label{k2e1jmfve!!!!2vec}
    \end{align}
     \end{subequations}
     }
\end{theorem}

 \begin{remark}
 {\rm
    The bound~\eqref{k2e1jmfve!!!!3} for scalar systems, which is \emph{independent} of the control policy and state value, has been developed using the concentration bounds of~\cite{rantzer2018concentration} for the scalar LS algorithm~\eqref{learningAlgorithm}. To the best of our knowledge, there are no similar concentration bounds for the vector variant of the LS algorithm~\eqref{learningAlgorithm} which work for \emph{any} 
    $\texttt{A}$, and a large class of control policies. Looking for such bounds, \blue{which are independent of the state value}, seems an interesting research venue. \blue{The lower bound~\eqref{k2e1jmfve!!!!2vec} is similar to~\eqref{k2e1jmfve!!!!2}, while~\eqref{k2e1jmfve!!!!2} is stronger for the particular case of scalar system, as the upper bound on the estimation error derived in ~\lemref{lem:Raginskyerror} is not required for the scalar case, and the estimation error is given in~\eqref{estsimerr222336227}.}
   } \oprocend
\end{remark}
 \blue{
\begin{exmp}
\label{examp1vect}
{\rm 
\begin{figure}[t]
\centering      
    \includegraphics[width = \columnwidth, trim = {0 0 0 2\baselineskip}]
    {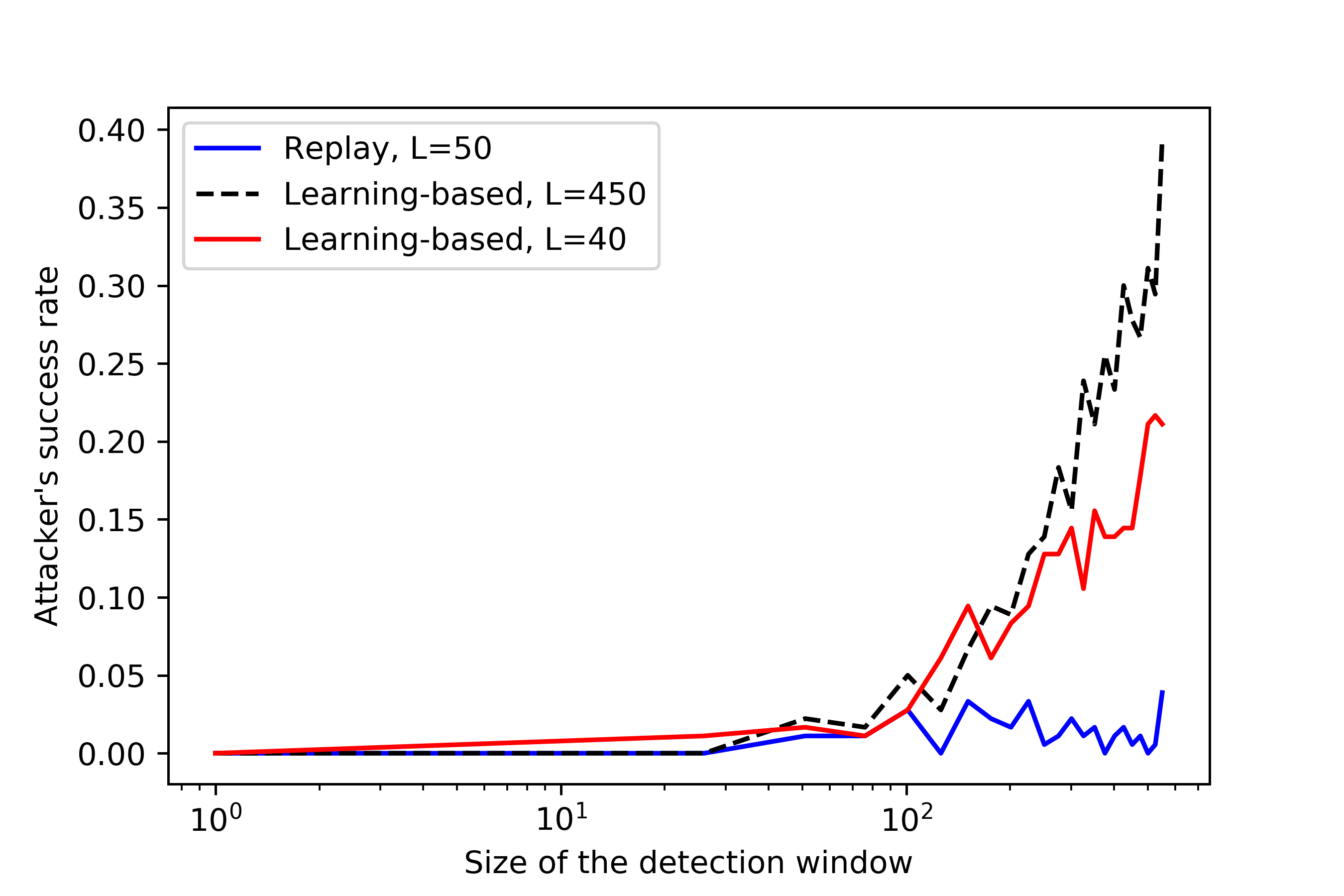}
        \caption{The attacker's success rate $\Pdec^{\texttt{A},T}$ versus the size of the detection window $T$.}
  \label{fig:mnonlinr34959}
  \end{figure}
In this example, we compare the empirical performance of the covariance test against the learning-based attack which utilizes LS estimation~\eqref{learningAlgorithmvect-var}, and the replay attack.  
At every time $k$,
the controller
tests the empirical covariance for anomalies over a detection window  $[1, T]$,
using a confidence interval $2 \gamma > 0$ around the operator norm of error matrix $\bDelta$~\eqref{Testvec}.
Since we are considering the 
Euclidean norm for vectors, the induced operator norm amounts to $\|\bDelta\|_{op}=\sqrt{\lambda_{max}(\bDelta^{\dagger}\bDelta)}$. 
Here, $\gamma=0.1$, $\bU_k = -0.9\texttt{A}\bY_k$ for all $1\le k\le T=600$, 
\begin{align}
\label{eq:ex:no-watermark:params}
    \texttt{A} &= \begin{pmatrix} 1 & 2 \\ 3 & 4 \end{pmatrix},
    &\bSigma   &= \begin{pmatrix} 1 & 0 \\ 0 & 2 \end{pmatrix}.
\end{align}

\figref{fig:mnonlinr34959} presents the performance averaged over $180$ runs of a Monte Carlo simulation.
It illustrates that the vector variant of our learning-based attack also \textit{outperforms} the replay attack. A learning-based attack with a learning phase of length $L=40$ has a higher success rate than a replay attack with a larger recording length of $L=50$. Similarly to the discussion for scalar systems in~\secref{perf2039943!}, the false-alarm rate decays to zero as the size of the detection window $T$ tends to infinity. Thus, the success rate of learning-based attacks increases as the size of the detection window increases. 
Finally, as illustrated in~\figref{fig:mnonlinr34959}, the attacker's success rate increases as the duration of the learning phase $L$ increases, since the attacker improves its estimate of $\texttt{A}$ as $L$ increases.
}\oprocend
\end{exmp}
An example that investigates the effect of  privacy-enhancing signals on the empirical deception probability of the learning-based attacks on the vector systems 
is in~\appref{oeke34-44canbefoxamplenumLQR44}.
}




\section{Discussion and Future Work}
\label{sec:conc2}


\subsection{Upper Bound on the Deception Probability}

The upper bound in \eqref{implici-vec3}, which relates the deception criterion \eqref{feji-vec3} to the estimation error $\|\bhA-\texttt{A}\|_{op}$, is used to find the lower bound \eqref{k!!vec!}. Finding a corresponding lower bound in term of $\|\bhA-\texttt{A}\|_{op}$ for~\eqref{implici-vec3:basic} is the first step   in extending our results in \thmref{conversew} to vector systems. Finding an upper bound on the deception probability for vector linear and scalar nonlinear systems, where the attacker can use \textit{any} learning algorithm, is  left open for future work.

 
\subsection{Model-based vs. Model-free}
\label{ss:basedvsfree}
In this work,  we mainly concentrated on linear systems; 
we assumed the attacker constructs the fictitious sensor reading, in a model-based manner, according to the linear model~\eqref{mimicmodel} and its vector variants. In general, as discussed in \appref{futrelatef2}, the system can be nonlinear, and the attacker might not be aware of the linearity or non-linearity of the dynamics.
Comparing the deception probability for the vast range of model-free and model-based
learning methods~\cite{tu2017leastrecht,Dean2019,satorras2019combining} is an interesting research venue. 


\subsection{Continuous Learning and Hijacking}
\label{ss:expbrffexploih}

In this work, we assumed two disjoint consecutive phases (recall \secref{ss:model:integrity_attack}): learning and hijacking, 
which are akin to the exploration and exploitation phases of reinforcement learning (RL) \cite{sutton1998reinforcement}.
Indeed, in this two-phase process, the attacker explores the system until it reaches a desired deception probability and then moves to the exploitation phase during which it drives the system to instability as quickly as it can. 
The two phases are completely separate  due to the inherent tension between them: exploiting the system without properly exploring it during the learning (silent) phase  increases the chances of being detected. 

Despite the naturalness of two-phase attacks, 
just like in RL~\cite{sutton1998reinforcement}, 
one may consider more general strategies where exploration and exploitation are intertwined and gradual: 
as time elapses, the attacker can gain better estimates of the observed system and \textit{gradually} increase its \textit{detection-limited} attack.
In these terms, our two-phase attack can be regarded as a two-stage approximation of the gradual attack and provides achievability bounds for such attacks.
Studying more general attacks is a  research venue that  is left for future study.

\rem{
However, more generally, the attacker can also hijack the sensor and control signals during the learning phase (see \figref{fig:sns}). 
However, as during the learning phase, the actual sensor reading are observed by the controller, the attacker's hijacking capabilities are restricted by the detector.
We have shown that an attacker that first passively learns the plant dynamics and then intervenes as a MITM, can completely destroy the plant while acting as a legitimate plant for the controller to remain undetected. 
We also assumed that 
during the hijacking phase (see \figref{fig:sns}), 
the attacker chooses its malicious control inputs to destroy the plant, as quickly as possible. 
Hence, it is reasonable to assume that the attacker uses a fixed estimate of the open-loop gain throughout the entire hijacking phase. 
Nevertheless, the attacker could devote some of its control inputs to refine its estimate of the open-loop gain during the hijacking phase. 
This is reminiscent to exploration--exploitation dilemma in  reinforcement learning~\cite{sutton1998reinforcement} where exploration corresponds to learning the plant and reduce the detection probability, and exploitation corresponds to destruction of the plant. Nonetheless, assuming that the attacker estimation can only improve during the hijacking phase, \thmref{LB:dec-prob} for scalar systems and~\thmref{LB:dec-prob-vec} for vector systems, provide lower bounds for this refined continuous-learning attack. The detailed analysis of a learning-based attacker, which simultaneously learns and hijacks in both phases, is an exciting research venue.

}

\subsection{Oblivious Controller}
\label{ss:learning_controller}

A more realistic scenario is the one in which neither the attacker nor the controller are aware of the open-loop gain of the plant. 
In this scenario, both parties strive to learn the open-loop gain---posing two conflicting objectives to the controller, who, on the one hand, wishes to speed up its own learning process, 
while, on the other hand, wants to slow down the learning process of the attacker.

In such a situation, standard adaptive control methods are clearly insufficient, 
as no asymmetry between the attacker and the controller can be achieved under the setting of~\secref{s:model}.
To create a security leverage over the attacker, 
the controller needs to utilize a judicious privacy-enhancing signal: 
A properly designed privacy-enhancing signal should enjoy a positive double effect by facilitating the learning process of the controller while hindering that of the attacker at the same time. Note that, in such a scenario, while the controller knows both $\oU_k$ and $\Gamma_k$, the attacker is cognizant of only their sum~\eqref{eq:watermaked_control}---$U_k$. This is
reminiscent of strategic information transfer~\cite{akyol2015privacy}.

Finally, we note that, unless the controller is able to detect an MITM attack (the attacker's hijacking phase), 
its learning process will be hampered by the fictitious signal that is generated according to the virtual system of the attacker~\eqref{mimicmodel}.
 
\subsection{Moving Target Defense}
\label{enriij3543}
In our setup the attacker has
full access to the control signal (see \figref{fig:sns}), and at time $k+1$ the attacker uses the control input $U_k$ to construct the fictitious sensor reading $V_{k+1}$ according to~\eqref{mimicmodel}. Thus, the \textit{watermarking} signal~\cite{MoSinopoli:Magzine}, the private random signature, might not be  an effective
way to counter the learning based attacks \blue{[cf.~Remark 9]}. Here, we introduced the \textit{privacy-enhancing} signal~\eqref{eq:watermaked_control} to impede the learning
process of the attacker and decrease the deception probability. 
Another technique that has been developed in the literature to counter attacks, where the attacker has full system knowledge, is having the controller covertly introduce virtual state variables unknown to the attacker but that are correlated with the ordinary state variables, so that modification of the original states will impact the extraneous states.  These extraneous states can act as a \emph{moving target}~\cite{weerakkody2015detecting,kanellopoulos2019moving,zhang2019analysis,griffioen2019optimal} for the attacker. A  similar technique is the so-called
\emph{baiting}, which adds an offset to the system dynamic~\cite{ flamholz2019baiting,hoehn2016detection}. In practice, this technique breaks the information symmetry between the attacker---which has the full system knowledge, and the controller.
Using such defense techniques to hamper the learning process of our proposed attacker, is an interesting research venue. In this context, the controller,  by potentially sacrificing the optimally of the control task, can act in an
adversarial learning setting. Assuming that the control can covertly introduce a virtual part to the dynamics, for any given duration of the learning phase $L$ (see \figref{fig:sns}),   sufficiently fast changes in the cipher part of the dynamic can drastically hamper the learning process of the attacker. Also,  as discussed in \appref{futrelatef2}, adding a rich nonlinearity to the dynamics can be used as a way to secure the system against learning-based attacks.
\blue{
\subsection{Optimal testing}
\label{ss:Testing_at_all_time}

	Throughout this work, we have assumed that the controller tests the integrity of the system at a specific time step $T$, that tends to infinity. 
    Since the controller does not know the exact time instant at which an attack might occur, a more realistic scenario would be that of continuous testing, i.e., that in which the integrity of the system is tested at every time step and where the false alarm and deception probabilities are defined with a union over time. We leave this treatment for future research. 
    
    In addition,  following~\cite{Kumar:DynamicWatermarki}, we have considered the variance-based test, which searches for anomalies in the empirical
variance, i.e., whether it falls outside a
confidence interval of length $2\delta$ [cf.~\eqref{Test1}]. Studying the optimal detector for learning-based attacks is an interesting research venue.
}
\subsection{Further Future Directions}

Other future directions can explore the extension of the established results to \blue{partially-observable linear vector systems where the input (actuation) gain is unknown}, characterising securable and unsecurable subspaces~\cite{satchidanandan2018control} for learning-based attacks, revising the attacker full access to both sensor and control signals, designing optimal privacy-enhancing signals (recall Remark~\ref{rem405322!}) for linear and nonlinear systems, \blue{investigating the scenario in which the attacker is  oblivious of the noise covariance matrix or more generally the noise statistics, and studying the relation between our proposed \textit{privacy-enhancing} signal with the noise signal utilized to achieve \textit{differential privacy}~\cite{cortes2016differential}}. 




\bibliographystyle{IEEEtran}
\bibliography{main}


\ver{
\vspace*{-3\baselineskip}
\begin{IEEEbiography}
[{\includegraphics[height=1.25in,clip]{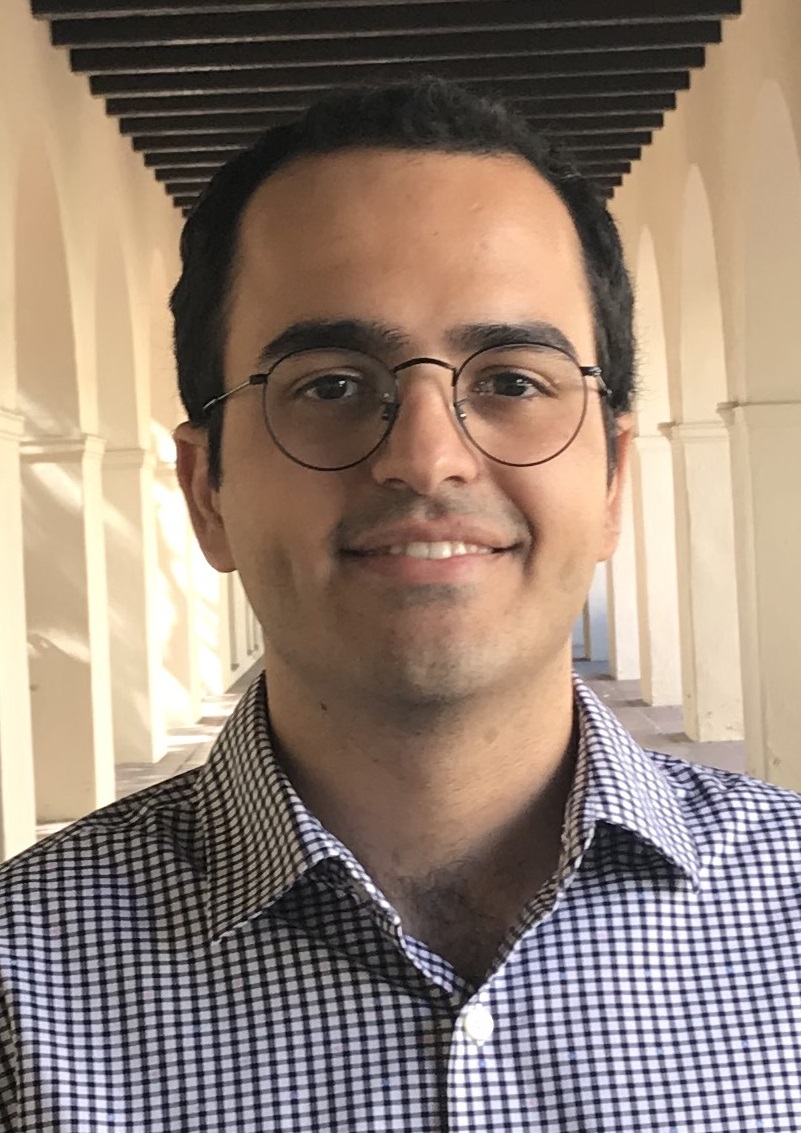}}]{Mohammad
    Javad Khojasteh}(S'14) did his undergraduate studies at  Sharif University of Technology from which he received double-major B.Sc.\ degrees in Electrical Engineering and in Pure Mathematics, in 2015. 
  He received the M.Sc.\ and Ph.D.\ degrees in Electrical and Computer Engineering from  University of California San Diego (UCSD), La Jolla, CA, in
  2017, and 2019,
respectively. Currently, he is a Postdoctoral Scholar in 
Center for Autonomous Systems and Technologies (CAST)
at California Institute of Technology,
  Pasadena, CA.
\end{IEEEbiography}
\vspace{-2\baselineskip}

\begin{IEEEbiography}[{\includegraphics[height=1.25in,clip]{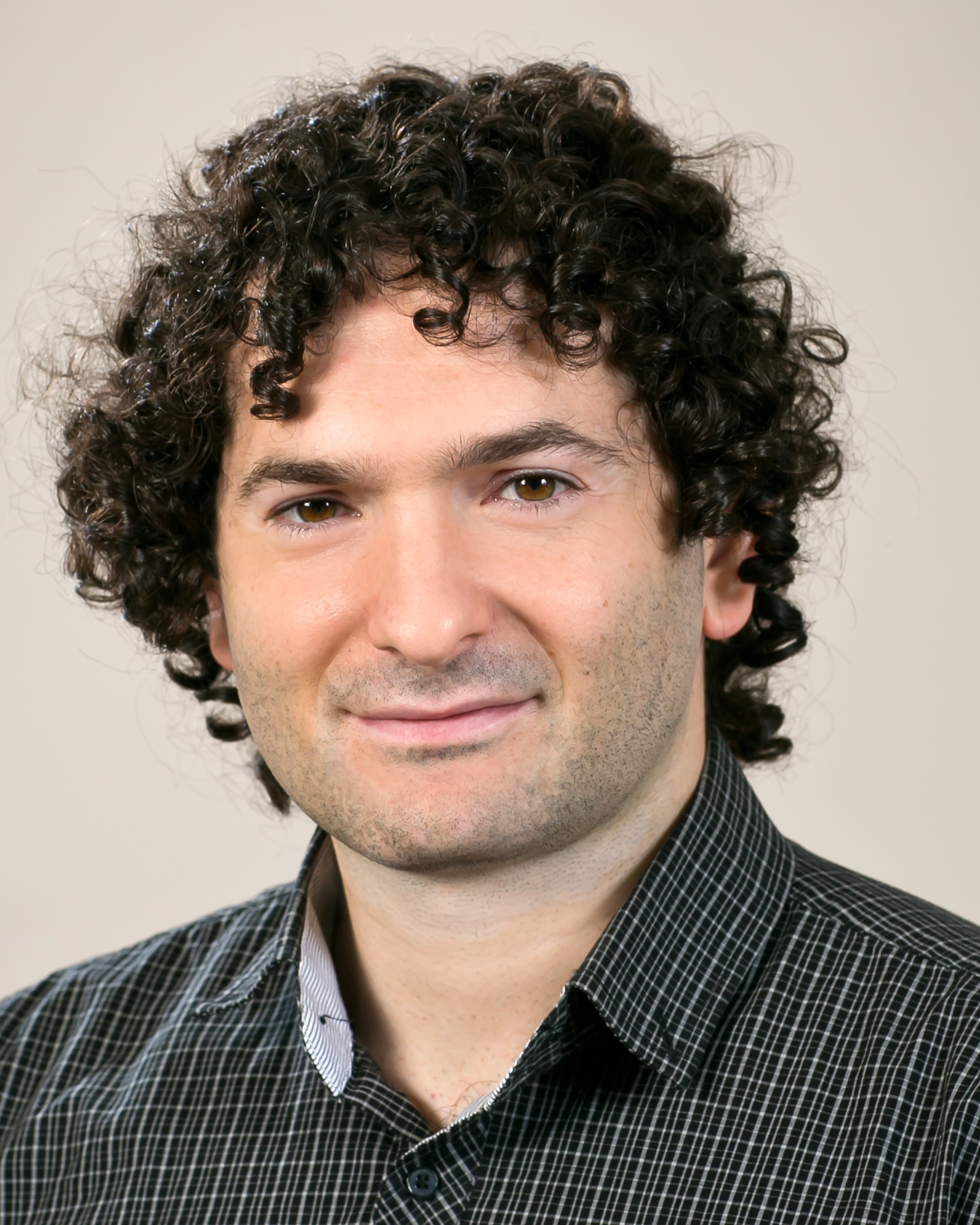}}]{Anatoly Khina} 
	(S'08--M'17)
    is 
    a Senior Lecturer in the 
    School of Electrical Engineering, Tel Aviv University, 
    from which he holds B.Sc.\ (2006),
    M.Sc.\ (2010),
    and Ph.D.\ (2016) degrees, all in Electrical Engineering.    
    He was a Postdoctoral Scholar in the Department of Electrical Engineering, California Institute of Technology,
    from 2015 to 2018, and a Research Fellow at the Simons Institute for the Theory of Computing, University of California, Berkeley, 
    during the Spring of 2018. 
%
\end{IEEEbiography}
\vspace*{-2.9\baselineskip}

\begin{IEEEbiography}
  [{\includegraphics[height=1.25in,clip]{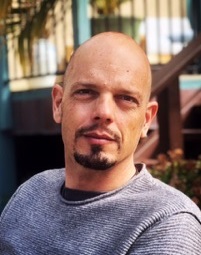}}]
  {Massimo Franceschetti} (M'98--SM'11--F'18) received the Laurea degree
  (with highest honors) in computer engineering from the University  Federico II, Naples, Italy, in 1997, the M.S.\ and Ph.D.\ degrees in
  electrical engineering from the California Institute of Technology,
  in 1999, and 2003, respectively.  He is Professor of
  Electrical and Computer Engineering at the University of California
  at San Diego (UCSD). Before joining UCSD, he was a postdoctoral
  scholar at the University of California at Berkeley for two
  years. He has held visiting positions at the Vrije Universiteit
  Amsterdam, the \'{E}cole Polytechnique F\'{e}d\'{e}rale de Lausanne,
  and the University of Trento. His research interests are in physical
  and information-based foundations of communication and control
  systems. 
   He was awarded the C. H. Wilts
   Prize in 2003 for best doctoral thesis in electrical engineering at
   Caltech; the S.A.  Schelkunoff Award in 2005 for best paper in the
   IEEE Transactions on Antennas and Propagation, a National Science
   Foundation (NSF) CAREER award in 2006, an Office of Naval Research
  (ONR) Young Investigator Award in 2007, the IEEE Communications
   Society Best Tutorial Paper Award in 2010, and the IEEE Control
   theory society Ruberti young researcher award in 2012. 
  He has been elected fellow of the IEEE in 2018 and became a Guggenheim fellow for the natural sciences, engineering, in 2019.
\end{IEEEbiography}
\vspace{-2\baselineskip}
\begin{IEEEbiography}
[{\includegraphics[width=1in,height=1.25in,clip,keepaspectratio]{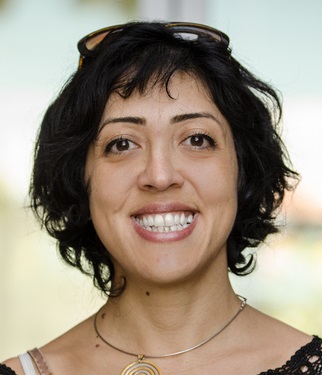}}]
{Tara Javidi}(S'-96--M'02) studied electrical engineering at Sharif University of Technology, 
from 1992 to 1996. She received her MS degrees in
electrical engineering (systems), and in applied mathematics (stochastics)
from the University of Michigan, Ann Arbor. She received her Ph.D.\ in
electrical engineering and computer science from the University of Michigan,
Ann Arbor, in May 2002.
From 2002 to 2004, Tara was an assistant professor of electrical engineering at the University of Washington, Seattle. She joined University of
California, San Diego, in 2005, where she is currently a professor of electrical
and computer engineering. 
She is a member of the Information Theory Society Board of Governors and
a Distinguished Lecturer of the IEEE Information Theory Society (2017/18).
Tara Javidi is a founding co-director of the Center for Machine-Integrated Computing and Security, a founding faculty member of Halicioglu Data Science Institute (HDSI) at UCSD.
\end{IEEEbiography}
}{} 

\newpage
\appendices

\newpage
\onecolumn
\newpage
\twocolumn

\section{Extension to Nonlinear Systems}
\label{futrelatef2}

In the previous sections, we focused on linear systems, 
where, for finding a lower bound on the deception probability of the learning-based attack, the LS algorithm has been utilized. 
In this section, we extend this treatment to nonlinear systems.

Finding the \textit{optimal} solution of nonlinear LS problems is difficult in general.
Consequently, different approaches to this problem have been introduced.
Prominent now-classical solutions include the Gauss--Newton and Levenberg--Marquardt algorithms \cite{marquardt1963algorithm}.
For richer nonlinear dynamics families, more sophisticated learning algorithms have been recently proposed, such as Gaussian processes (GP) regression~\cite{deisenroth2011pilco,berkenkamp2017safe,umlauft2017feedback,srinivas2012information,shekhar2018gaussian,khojasteh2019probabilistic22} and deep neural networks~\cite{gal2016improving,chen2018approximating}; in this section, we concentrate on GP inference.

To derive analytical results, we need to restrict the class of possible dynamics. 
Similarly to our treatment for linear systems, we follow the frequentist approach.\footnote{We assume the attacker uses GP regression to \textit{construct the learning algorithm} despite the function $f$ governing the system dynamics assumed to be non-random (cf.~\cite{chowdhury2017kernelized}).}
We work with the natural counterpart of GP---the Reproducing Kernel Hilbert Space (RKHS)~\cite{berkenkamp2017safe,chowdhury2017kernelized}. 
To that end, we consider the following scalar nonlinear system
\begin{align}
\label{nonlinearsystem1}
   X_{k+1} = f(Z_k) + W_k,
\end{align}
where $Z_k=(X_k,U_k) \in \real \times \cU \subseteq \real^2$; the plant disturbance process $\{W_k\}$ has i.i.d.\ Gaussian samples of zero mean and variance $\sigma^2$; $f: \mathcal{X} \times \mathcal{U} \rightarrow \real$ belongs to the class of RKHS functions spanned by the symmetric bilinear positive-definite kernel $\mathcal{K}(Z,Z')$ that specifies the RKHS. The inner product of the RKHS satisfies the reproducing property: $f(Z)=\inner{f}{\mathcal{K}(Z,.)}_\cK$.
The choice of the kernel is problem and application dependent; see, e.g., \cite{williams2006gaussian} for a review of common kernel choices.
The RKHS norm, induced by the kernel $\mathcal{K}$, is defined as $\|f\|_\mathcal{K} \triangleq \sqrt{\inner{f}{f}_\cK}$,
that can be viewed as a complexity measure of the function $f$ w.r.t.\ the kernel $\mathcal{K}$. We assume $\|f\|_\mathcal{K} \le \chi$ for a known constant $\chi > 0$.
The function $f$ is assumed to be completely known to the controller, while the attacker is oblivious of $f$ but is assumed to know the RKHS to which $f$ belongs.

As in~\eqref{L2Lshouldsatf}, under legitimate system operation, the controller observation $Y_k$ behaves according to 
\begin{align}
\label{L2Lshoulnonlineaf34}
	Y_{k+1} -  f(Y_k,U_k)\sim ~\iid~ \mathcal{N}(0,\sigma^{2}).
\end{align}
Thus, following the exposition of \secref{ss:model:variance-test}, 
we consider the variance test, i.e., the controller tests whether the empirical variance of \eqref{L2Lshoulnonlineaf34} falls within a confidence interval of length $2 \delta > 0$ around its expected variance $\sigma^2$ [cf.\ \eqref{Test1}]. That is, at test time $T$, it checks whether
\begin{align}
\label{test4nonlinear443}
\begin{aligned}
	&\frac{1}{T} \sum_{k=1}^{T} \left[ Y_{k+1}-f(Y_k,U_k) \right]^2 
 \\ &\qquad\qquad\qquad 
    \in (\Var{W}-\delta, \Var{W}+\delta).
\end{aligned}
\end{align}

Denote by $\hat{F}$ the estimation of the attacker of the function $f$ at the conclusion of Phase 1. 
We  assume that during Phase 2, 
the fictitious sensor reading
is constructed in a model-based fashion according to the nonlinear analogue of~\eqref{mimicmodel}:
\begin{align}
\label{mimicmodelnonoe33}
	V_{k+1} &= \hat{F} (V_k, U_k) + \tW_k \,, & k = L, \ldots, T-1 .
\end{align}
%

We now extend the results of \secref{sec:successful} to our nonlinear setup.
To provide a lower bound on the deception probability $\Pdec^{f,T}$, we assume that the attacker uses GP inference, 
which is known to be closely related to RKHS functions (cf.~\cite{berkenkamp2017safe,chowdhury2017kernelized}), 
to construct an estimate $\hF$ of $f$ at the conclusion of the learning phase. 
To that end, the function $f$ is assumed to have a Gaussian prior 
and treat it as a GP, $F$, with a kernel (covariance function) $\cK$, which is the kernel associated with the RKHS to which $f$~\eqref{nonlinearsystem1} belongs. Without loss of generality, we assume $F$ to be of zero mean.

\begin{remark}
Following~\cite{chowdhury2017kernelized}    the assumption that $f$ is random and  distributed according to a GP is used only for the attacker's learning algorithm construction and is not part of the model setup \eqref{nonlinearsystem1}.
\end{remark}

During the learning phase, the attacker observes 
the state-and-control trajectories $Z_1^{L-1}$
up to time $L-1$. 
That is, the attacker observes $Z_1^{L-1}$---the $(L-1)$ inputs to the function $F$, and $X_2^{L}$---their corresponding outputs corrupted by the i.i.d.\ Gaussian system disturbances $W_1^L$.
Then, for all $1 \le k\le L-1$, the posterior distribution of $F$ is also a GP with mean $\mathcal{M}_k$, covariance $\texttt{K}_k$, and variance $\sigma_k^2$, given by (see, e.g., \cite{williams2006gaussian}):
\begin{align}
\label{GPalgoriti345666}
\begin{aligned}
    \mathcal{M}_k(Z)&=\mathcal{C}_k(Z)(\bar{\mathcal{C}}_k+ \Id_{k \times k} \sigma^2)^{-1} (X_2^{k+1})^\dagger,
    \\
    \!\!\!\!\!\!
    \texttt{K}_k(Z,Z')&=\mathcal{K}(Z,Z')-
    \mathcal{C}_k(Z)(\bar{\mathcal{C}}_k+ \Id_{k \times k} \sigma^2)^{-1}\mathcal{C}^\dagger_k(Z'), \:\:
    \\
    \sigma_k^2(Z)&=\texttt{K}_k(Z,Z),
\end{aligned}
\end{align}
where the vector $\mathcal{C}_k=[\mathcal{K}(Z,Z_1) \ldots \mathcal{K}(Z,Z_k)]$ comprises the covariances between the input $Z$ and the observed state-and-control trajectories vector $(Z_1^k)^\dagger$, and  $\bar{\mathcal{C}}_k \in \real^{k \times k}$ is the covariance matrix of the vector $(Z_1^k)^\dagger$, i.e., $(\bar{\mathcal{C}}_k)_{i,j}=\mathcal{K}(Z_i,Z_j)$.

Unlike the linear case, where the attacker needs to learn finitely many parameters, in the case of RKHS functions, in general,
there are
infinitely many parameters 
to learn.\footnote{\blue{In fact, having a ``rich" nonlinear dynamics can be used as a way to secure the system against learning-based attacks (cf.\ \secref{enriij3543}).  By ``rich dynamics" we mean that $f$ belongs to a space of functions that can be learned only by a statistical algorithm that has high expressive power or Vapnik--Chervonenkis (VC) dimension. } } In this case, to  have a positive asymptotic deception probability, the duration of the learning phase should be sufficiently large. Namely, in our analysis, we assume
\begin{align}
\label{nnnloinea4456relation45}
    T=L+c,
\end{align}
where $c \ge 1$ is a positive constant, and we investigate the deception probability in the limit of $L \to \infty$. 

By choosing $\hat{F}=\mathcal{M}_{L-1}$---the minimum mean square error estimator of $F$, 
we can bound from below the asymptotic deception probability under the variance test 
of the \textit{best} nonlinear learning-based attack~\eqref{mimicmodelnonoe33}
for \textit{any measurable} control policy, as follows (the proof is available in the appendix).

\begin{theorem}
\label{LB:dec-prob-nonlinear3}
    Given~\eqref{nnnloinea4456relation45}, the asymptotic deception probability  of the nonlinear learning-based attack~\eqref{mimicmodelnonoe33}, under the variance test~\eqref{test4nonlinear443}, is bounded from below by  
    \begin{align}
    \label{!nonlinearporbfer}
        \lim_{L \to \infty} \Pdec^{f,L+c} \ge \lim_{L \to \infty} \bar{p}\prod_{k=L+1}^{L+c} (1-\xi_k)
    \end{align}
    for all $f$ in the RKHS, where $0 \le \xi_k \le 1$ is defined as
    \begin{subequations}
    \begin{align}
    \label{akhr349595922!!}
        \xi_k &\triangleq e^{\psi_{L-1}+1-\left(\dfrac{\nu_k-\chi}{4\sigma\sigma_L(V_k,U_k)}\right)^2} ,
    \\
    \label{phiiii3494}
        \psi_{L-1} &\triangleq \frac{1}{2}\sum_{k=1}^{L-1}\ln(1+\sigma^{-2}\sigma_{k-1}^2(Z_k)),
    \end{align}
    \end{subequations}
    and $\{\nu_k\}$ is a sequence of non-negative reals  such that
    \begin{align}
    \label{iefi54076543!!!1}
       \lim_{L \rightarrow \infty} \frac{1}{L+c}\sum_{k=L+1}^{L+c} \nu_k^2 + \frac{2}{L+c} \sum_{k=L+1}^{L+c} |\tW_k|\nu_k \le \delta.
    \end{align}
    holds with probability $\bar{p}$.
\end{theorem}

  By looking at~\eqref{!nonlinearporbfer}, it follows that the bound on the deception probability depends on the variances of the posteriors at the observation points $Z_1^{L-1}$, as well as $\sigma_L(V_k,U_k)$ for $L+1 \le k \le L+c$. In particular, as the variances of posteriors are smaller, the attacker has less uncertainty about the system dynamics, and  its success rate increases. 

\begin{remark}
    Designing a proper privacy-enhancing signal which leads to large variances for posteriors, and consequently, enhanced detection probability is an interesting future research venue.  \oprocend  
\end{remark}

\blue{
\begin{exmp}
\label{ex:scalarnolin:deception-rate}
      {\rm
    In this example, we investigate the performance of attacks based on nonlinear Gaussian-processes (GP) learning algorithms. At  time $T$,
    the controller
    tests the empirical variance for anomalies  over a detection window  $[1, T]$,
    using a confidence interval $2 \delta > 0$ around the expected variance~\eqref{test4nonlinear443}. 
    %
    Here, $f(X_k,U_k)=X_k^2+\sin (X_k)+U_k$ in~\eqref{nonlinearsystem1},  $\delta=0.1$, $U_k = -1.1 X_k^2$ for all $1\le k\le T=400$, $\{W_k\}$ are i.i.d.\ standard Gaussian, and the learning-based attacker~\eqref{mimicmodelnonoe33} uses the GP algorithm~\eqref{GPalgoriti345666}, with  sum-kernel including Radial-basis function (RBF) kernel and a White kernel~\cite[Ch.~4]{{williams2006gaussian}}, to estimate $f$. 
   
    \figref{fig:mnahanonie} presents the performance averaged over $350$ runs of a Monte Carlo simulation.
 Similarly to the discussion for linear systems in~\secref{perf2039943!}, the false-alarm rate decays to zero as the size of the detection window $T$ tends to infinity. Thus, the success rate of learning-based attacks increases as the size of the detection window increases. 
      As illustrated in~\figref{fig:mnahanonie}, 
\begin{figure}[t]
\centering      
    \includegraphics[width = \columnwidth, trim = {0 0 0 2\baselineskip},clip]
    {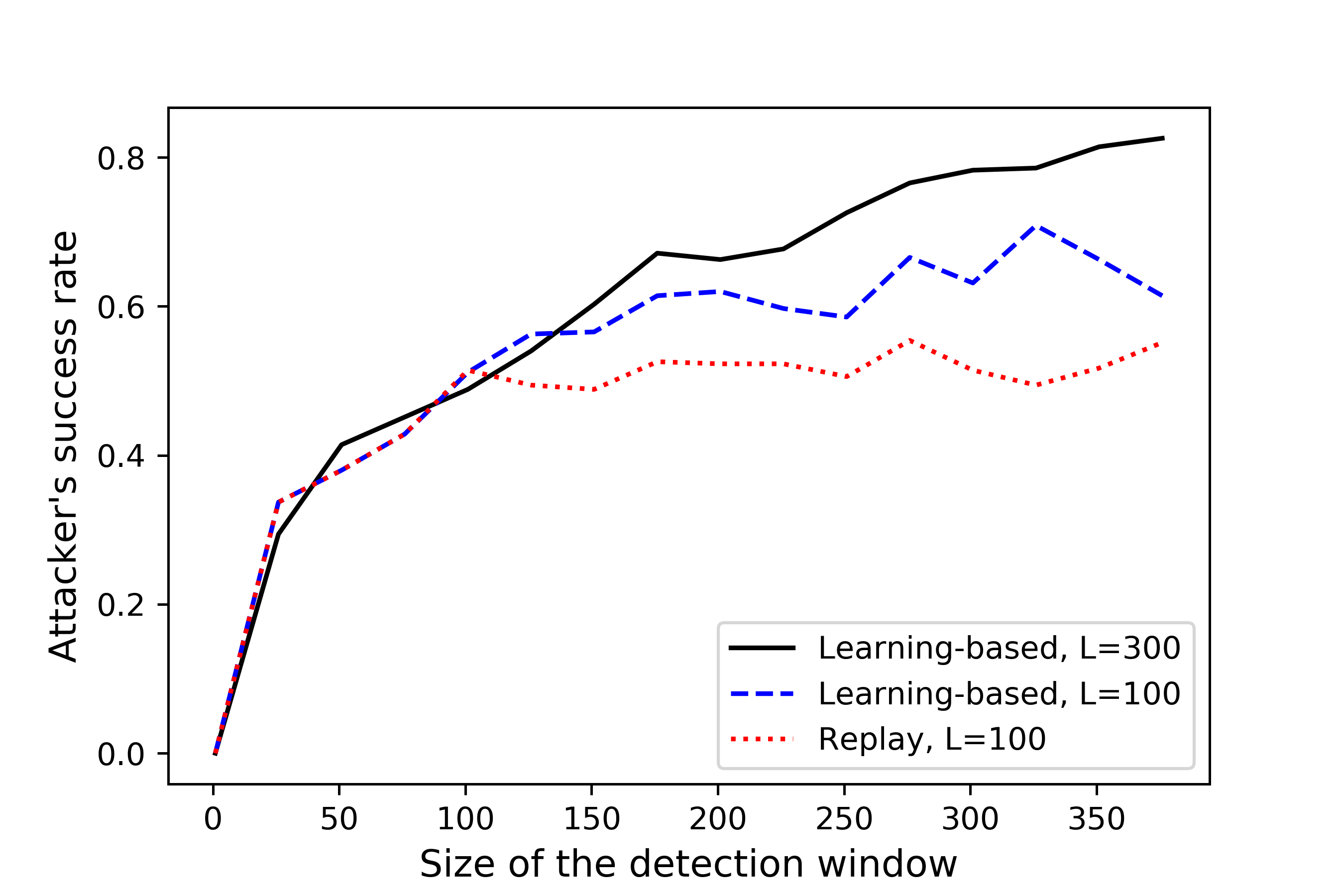}
        \caption{The attacker's success rate $\Pdec^{a,T}$ versus the size of the detection window $T$. 
    }
  \label{fig:mnahanonie}
\end{figure}
    the attacker's success rate increases as the duration of the learning phase $L$ increases. This is in agreement with~\eqref{!nonlinearporbfer} since the attacker can improve its posterior and the variance $\sigma_L(V_k,U_k)$ reduces as $L$ increases. 
   This figure also illustrates that the nonlinear variant of our learning-based attack \textit{outperforms} the replay attack: A learning-based attack with a learning phase of length $L=100$ has a higher success rate than a replay attack with a recording length of $L=100$.}
    \oprocend
\end{exmp}
}

\blue{
\section{Additional Numerical Examples for Privacy-enhancing Signals}
\label{ExamplenumLQR44}

In this section, we provide additional numerical examples to illustrate the effect of privacy-enhancing signals of \secref{subsectionauthernt11}. 

\subsection{The authentication policy proposed in Example~\ref{re:rr444exap13}}
\label{ExamplenumLQR44111}
In Example~\ref{re:rr444exap13}, we introduced a class of privacy-enhancing signals $\Gamma_k$ that ensure enhanced guarantees on the detection probability when the attacker may use \textit{any} learning algorithm to estimate the open-loop gain $A$ that is uniformly distributed over the symmetric interval $[-R, R]$, at the expense of a degradation in the control cost.
Here, we numerically study the effect of this authentication policy on the (time-averaged) linear-quadratic (LQ) control cost~\cite{bertsekas2019reinforcement}: 
\begin{align}
\label{controlcost}
	\oJ_T(U) \triangleq \frac{1}{T}\Ep{\sum_{k=0}^{T} qX_k^2+rU_k^2},
\end{align}
where the weights $q$ and $r$ are non-negative real numbers 
that penalize the cost for state deviations and control actuations, respectively, and are known at the controller. In fact, we show that the class of privacy-enhancing signals described in Example~\ref{re:rr444exap13} yields better guarantees on the deception probability 
at the expense of an increase in the 
control cost \eqref{controlcost}.

To that end, we simulated 300 Monte Carlo experiments for a system with $a=1$, $r=q=1$, and i.i.d. standard Gaussian variables $\{W_k\}$. 
Also, we consider the legitimate system operation where $Y_k=X_k$ for all time. 

In~\figref{fig:mnahaLQR22}, we
compare 
the LQ control cost
as a function of time
for three different control policies: 
I) unauthenticated control signal~$\oU_1^k=-0.5 aY_k$ for all $k$,  II) authenticated control signal~\eqref{eq:watermaked_control}, where $\Gamma_k$ are given in Example~\ref{re:rr444exap13} and $\eta=2$,  III) authenticated control signal~\eqref{eq:watermaked_control}, where $\Gamma_k$ are given in Example~\ref{re:rr444exap13} and $\eta=5$. 
\begin{figure}[t]
 \centering      
     \includegraphics[width=\columnwidth, trim = {0 0 0 2.6\baselineskip}, clip]
     {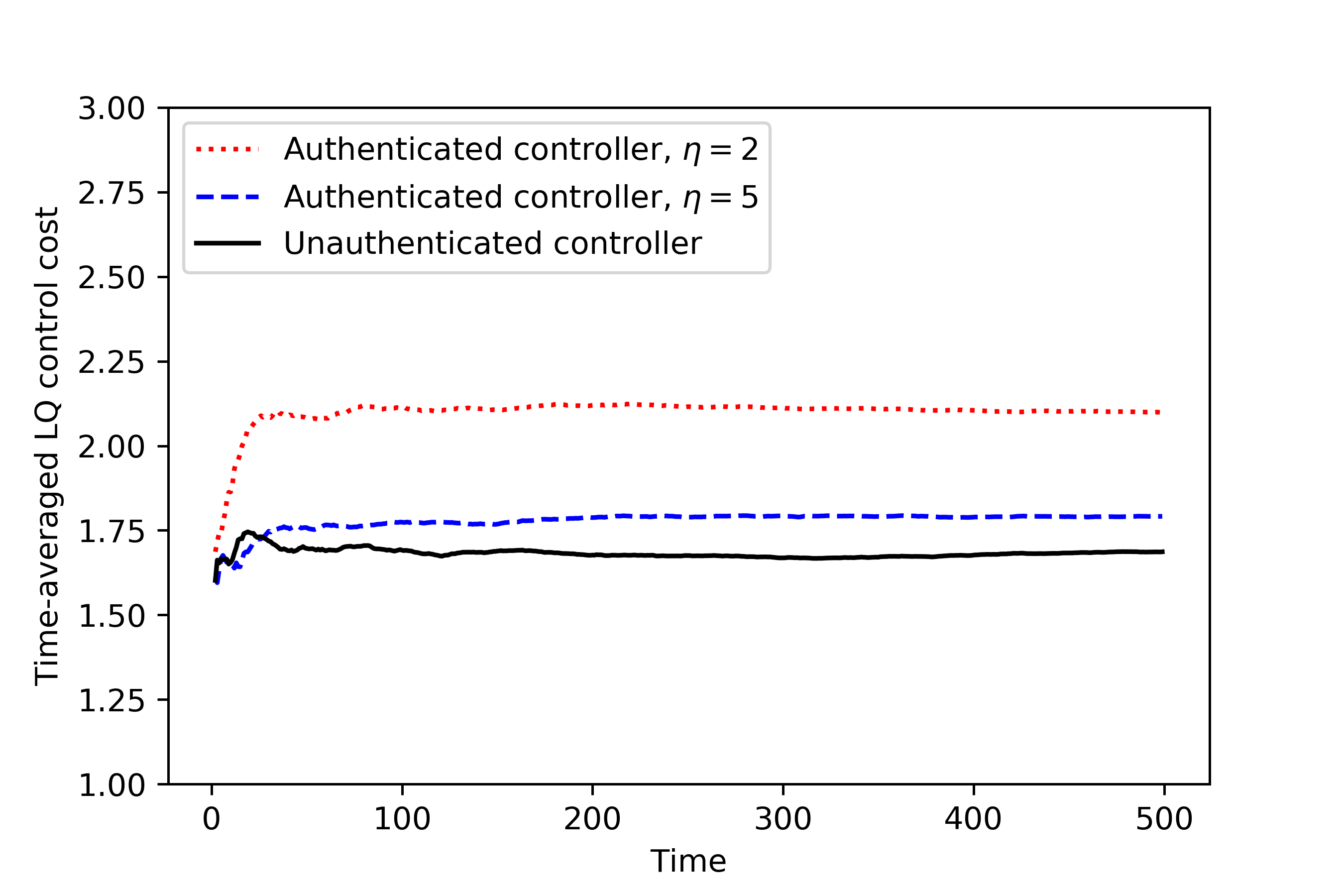}
         \caption{The (time-averaged) linear quadratic (LQ) control cost versus time.}
   \label{fig:mnahaLQR22}
\end{figure}
}

\begin{figure}[t]
 \centering      
     \includegraphics[width=\columnwidth, trim = {0 0 0 2.6\baselineskip}, clip]
     {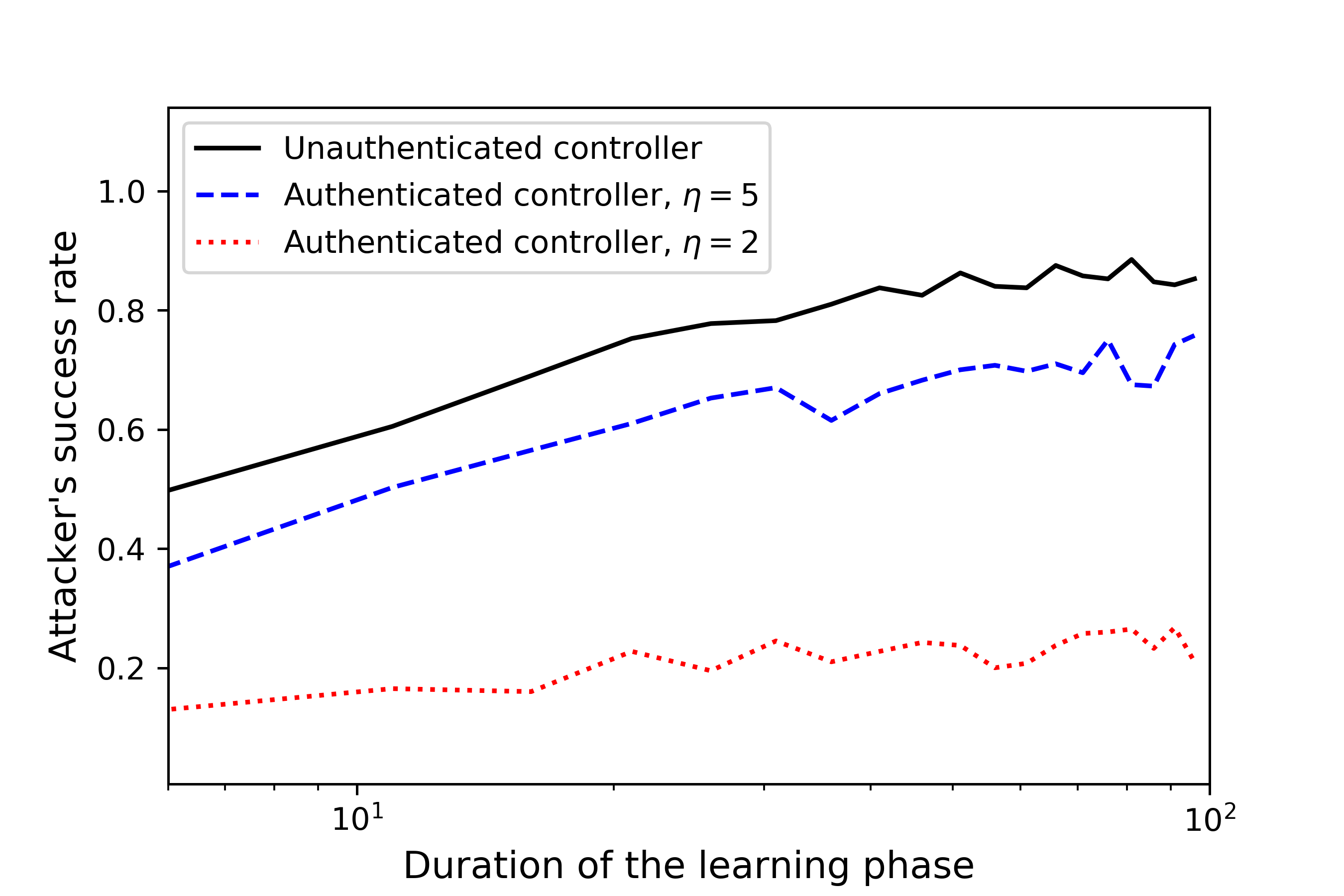}
         \caption{The attacker's success rate $\Pdec^{a,T}$ versus the duration of the learning phase $L$.  
         }
   \label{fig:WExpm4LSreis}
\end{figure}

\blue{
\figref{fig:WExpm4LSreis} illustrates the effect of the privacy-enhancing signal introduced in Example~\ref{re:rr444exap13} on the deception probability where the attacker particularly uses  the LS algorithm~\eqref{learningAlgorithm}.
  Here, the detector uses the variance test~\eqref{Test1}, $a=1$, $T=500$, $\delta=0.1$, and $\{W_k\}$ are i.i.d. standard Gaussian. We consider the same three control policies of~\figref{fig:mnahaLQR22}. $400$ Monte Carlo simulations were performed.

As discussed in Example~\ref{ex:authentication}, for the authenticated and unauthenticated control signals, the attacker's success rate increases as the duration of the learning phase increases.
Also, for a fixed $L$, the attacker performance deteriorates in the presence of an authenticated signal since $\Gamma_k$ hampers the learning process of the attacker.

Figs.~\ref{fig:mnahaLQR22} and~\ref{fig:WExpm4LSreis} demonstrate that the class of privacy-enhancing signals described in Example~\ref{re:rr444exap13} yields better guarantees on the deception probability 
at the expense of increasing the 
control cost \eqref{controlcost}.}

\blue{
\subsection{Vector Systems}
\label{oeke34-44canbefoxamplenumLQR44}

The following numerical example, which is the vector analogue of Example~\ref{ex:authentication}, illustrates the effect of the privacy-enhancing signal on the deception probability for the vector system.

}

\begin{figure}[t]
\centering      
    \includegraphics[width = \columnwidth, trim = {0 0 0 2\baselineskip}]
    {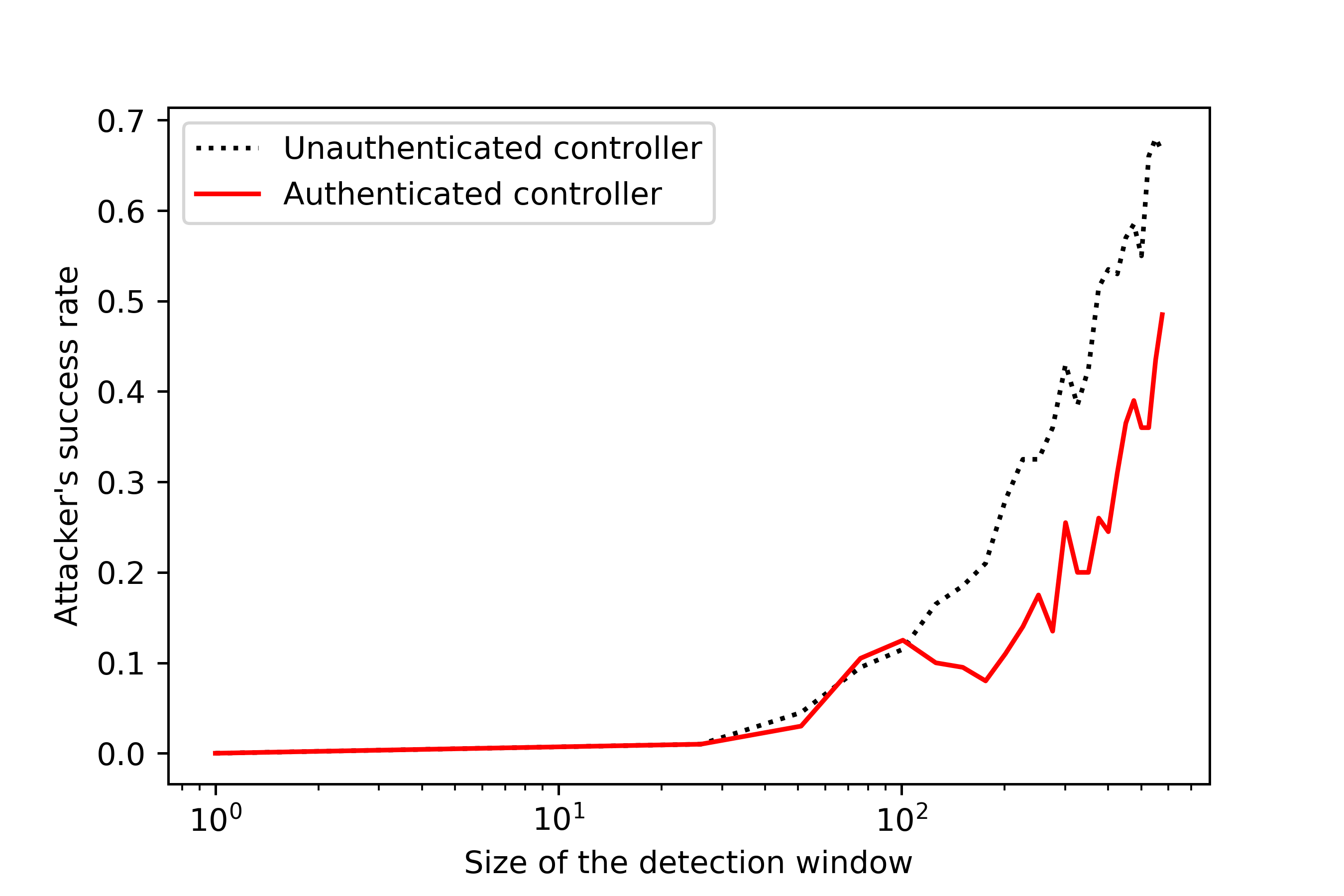}
        \caption{The attacker's success rate $\Pdec^{\texttt{A},T}$ versus the size of the detection window $T$.}
  \label{fig:mnonlinrnoonoo}
  \end{figure}

\begin{exmp}
\label{examp6ii33}
{\rm 
Consider the vector-plant setting with a privacy-enhancing signal (cf.~\secref{subsectionauthernt11}) 
and its yielded enhanced detection probability. 
As in Example~\ref{examp1vect}, we assume that the controller uses the empirical covariance test~\eqref{Testvec} and that the attacker utilizes LS estimation~\eqref{learningAlgorithmvect-var}. Again, the false alarm rate decays to zero as the detection window size $T$ goes to infinity. 
Here, $\gamma=0.1$,
$\texttt{A}$ is as in \eqref{eq:ex:no-watermark:params}, and $\bSigma = \texttt{I}_2$.
\figref{fig:mnonlinrnoonoo} compares 
the attacker's success rate, namely, the empirical $\Pdec^{\texttt{A},T}$,
as a function of the size of the detection window $T$
for two different control policies, averaged over 200 runs of a Monte Carlo simulation: 
I) Unauthenticated control $\bar{\textbf{u}}_1^k=-\texttt{A}\bY_k$ for all $1\le k\le T=600 $,  II) The vector analogue of the authenticated control signal of~\eqref{eq:watermaked_control}, where ${\Gamma}_k$ are i.i.d.\ zero-mean Gaussian 
with a diagonal covariance matrix with diagonal $\begin{pmatrix} 12, & 10 \end{pmatrix}$.
As is evident from \figref{fig:mnonlinrnoonoo}, the privacy-enhancing signal ${\Gamma}_k$ hampers the learning process of the attacker consequently reduces its deception probability.
}\oprocend
\end{exmp}


\section{\blue{More details for the proof} of Theorem~\ref{LB:dec-prob}}
\subsection{Proof of \lemref{Ineedtowk}}
\label{lemprofr111113}


    Since the hijacking phase of a learning-based attack~\eqref{mimicmodel} starts at time $k = L+1$, using~\eqref{eq:plant} and~\eqref{eq:subtract4test} we have
    \begin{subequations}
    \label{jf14053www}
    \noeqref{jf14053www:basic,jf14053www:explicit}
    \begin{align}
    \nonumber
    &\frac{1}{T} \sum_{k=1}^{T} (Y_{k+1}-a Y_k-U_k(Y_{1}^k))^2
    \\
    &= \frac{1}{T} \left(\sum_{k=1}^{L} W_k^2+ \sum_{k=L+1}^{T} (\tW_k+(\hA-a)V_k)^2\right)
    \label{jf14053www:basic}
    \\
    \nonumber
    &= \frac{1}{T} \left(\sum_{k=1}^{L} W_k^2+\sum_{k=L+1}^{T}\tW_k^2\right)
    \\
    &\quad + \frac{(\hA-a)^2}{T} \sum_{k=L+1}^{T}V_k^2+\frac{2(\hA-a)}{T} \sum_{k=L+1}^{T} \tW_k V_k.
    \label{jf14053www:explicit}
    \end{align}
    \end{subequations}
    Let $\mathcal{F}_k$ be the $\sigma-$filed generated by
    $\{(V_k,\hA,\tW_k,U_k)| k = L, \ldots, T - 1\}$. Then clearly, $V_{k+1}$ is $\mathcal{F}_k$ measurable, also $(\tilde{W}_{k+1},\mathcal{F}_k)$ is a martingale
difference sequence. Thus, using \cite[Lemma 2, part iii]{lai1982least} the last term in \eqref{jf14053www:explicit} 
    reduces to
    \begin{align}\label{eflko1133}
    \sum_{k=L+1}^{T} \tW_k V_k=o \left(\sum_{k=L+1}^{T} V_k^2 \right)+O(1) ~~~~\mbox{a.s.} 
    \end{align}
    in the limit $T \to \infty$.
    
    Note further that  
	\begin{align}
    \label{jjiief123331}
	    \lim_{T \rightarrow \infty}\frac{1}{T} \left(\sum_{k=1}^{L} W_k^2+\sum_{k=L+1}^{T}\tW_k^2\right)=\Var{W} \,~~\mbox{a.s.}~~~ 
    \end{align}
    by the strong law of large numbers~\cite{durrett2010probability}.
    
    Substituting \eqref{conversecss} in \eqref{jf14053www}, \eqref{jjiief123331}
    in \eqref{jf14053www}, using~\eqref{eflko1133}, and taking $T$ to infinity concludes the proof of the lemma.\hfill \, \QED
    \blue{
\subsection{More details for the proof of $(10a)$}
\label{lemprofr11111344}
	Under the variance test, 
    \begin{align}
    	\lim_{T\rightarrow \infty} \Pdec^{a,T}=\lim_{T\rightarrow \infty} \Ep{\mathbbm{1}_T},
    \end{align}
    where $\mathbbm{1}_T$ is one if~\eqref{Test1} occurs and zero otherwise. 
    Using the dominated convergence theorem~\cite{durrett2010probability} and \lemref{Ineedtowk}, we deduce 
    \begin{align}
    	\lim_{T\rightarrow \infty} \Pdec^{a,T}=\Ep{\mathbbm{1}'_T},
    \end{align}
     where $\mathbbm{1}'_T$ is one if 
        $(\hA-a)^2 / \beta \in(-\delta, \delta)$
     and zero otherwise,
    which concludes the proof.
}

\section{\blue{More details for the proof of} Theorem~\ref{conversew}}
\label{MMMwwie3444533333proofto}
\blue{
\subsection{More details for the proof of $(13a)$}}
\label{proof13as}
Using~\eqref{k2e1jmfve!!!!1} and \eqref{eq:det-prob:explicit} we deduce
\begin{align*}
     \lim_{T \rightarrow \infty}\Pdec^{T}&
 =\frac{1}{2R}\int_{-R}^{R}\mathbb{P}_a\left(|\hA-a|<\sqrt{\delta\beta}\right) da
 \\&=\frac{1}{2R}\int_{-R}^{R}\Ep{\mathbbm{1}_c} da,
\end{align*}
where $\mathbbm{1}_c$ is one if $|\hA-a|<\sqrt{\delta\beta}$ and zero otherwise. Consequently, using Tonelli's theorem ~\cite{durrett2010probability} it follows that
\begin{align}
\label{enien22!!!!!22}
     \lim_{T \rightarrow \infty}\Pdec^{T}=\mathbb{P}(|A-\hA|< \sqrt{\delta\beta}).
\end{align}
\subsection{Proof of (\ref{newlowreddw})}
\label{app:KL:manipulations}

	We start by applying the chain rule for mutual information to $\MI{A}{Z_1^L}$ as follows.
    \begin{align}
    	I(A;Z_1^L) &= \sum_{k=1}^L \CMI{A}{Z_k}{Z_1^{k-1}} .
    \label{eq:UB:proof:part2:chain-rule}
    \end{align}

	We next bound $\CMI{A}{Z_k}{Z_1^{k-1}}$ from above.
    \begin{subequations}
    \label{eq:UB:proof:part2}
	\begin{align}
    	& \CMI{A}{Z_k}{Z_1^{k-1}}
	 	= \CMI{A}{X_k, U_k}{Z_1^{k-1}}
    \label{eq:UB:proof:part2:explicitZ}
	 \\ &= \CMI{A}{X_k}{Z_1^{k-1}}  
    \label{eq:UB:proof:part2:markovity}
     \\ &= \CKL{\mathbb{P}_{X_k|Z_1^{k-1},A}}{\mathbb{P}_{X_k|Z_1^{k-1}}}{\mathbb{P}_{Z_1^{k-1},A}}
    \label{eq:UB:proof:part2:KL-MI}
     \\ &= \mathbb{E}_\mathbb{P}\left[\log\frac{d\mathbb{P}_{X_k|Z_1^{k-1},A}}
{d\mathbb{P}_{X_k|Z_1^{k-1}}}\right]
    \label{eq:UB:proof:part2:KL-def}
     \\ &= \mathbb{E}_\mathbb{P}\left[\log \frac{d\mathbb{P}_{X_k|Z_1^{k-1},A}}{d\mathbb{Q}_{X_k|Z_1^{k-1}}}\right] 
     -\mathbb{E}_\mathbb{P}\left[\log\frac{d\mathbb{P}_{X_k|Z_1^{k-1}}}{d\mathbb{Q}_{X_k|Z_1^{k-1}}}\right] \ \ 
    \label{eq:UB:proof:part2:measure-dominationh}
 \\ &\leq D(\mathbb{P}_{X_k|Z_1^{k-1},A}\| \mathbb{Q}_{X_k|Z_1^{k-1}}|\mathbb{P}_{Z_1^{k-1},A}) ,
    \label{eq:UB:proof:part2:KL-nonnegative}
    \end{align}
    \end{subequations}
    where we substitute the definition of $Z_k \triangleq (X_k, U_k)$ to arrive at \eqref{eq:UB:proof:part2:explicitZ}, 
    \eqref{eq:UB:proof:part2:markovity} follows from the chain rule for mutual information and the Markovity assumption $A\rightarrow  (X_k,Z_1^{k-1}) \rightarrow U_k$,
    we use the definition of the conditional mutual information in terms of the conditional KL divergence (recall the notation section) to attain \eqref{eq:UB:proof:part2:KL-MI} and \eqref{eq:UB:proof:part2:KL-def}, 
	the manipulation in \eqref{eq:UB:proof:part2:measure-dominationh} is valid due to the condition $\mathbb{P}_{X_k|Z_1^{k-1}} \ll \mathbb{Q}_{X_k|Z_1^{k-1}}$ in the setup of the lemma,  
	and \eqref{eq:UB:proof:part2:KL-nonnegative} follows from the non-negativity property of the KL divergence.

	Substituting \eqref{eq:UB:proof:part2} in \eqref{eq:UB:proof:part2:chain-rule} concludes the proof. \hfill \, \QED


\section{Proof of Corollary~\ref{uplowthem}}
Set $\mathbb{Q}_{X_k|Z_1^{k-1}}\sim \mathcal{N}(0,\sigma^{2})$. 
Then, $\mathbb{P}_{X_k|Z_1^{k-1},A} = \mathcal{N}(AX_{k-1}+U_{k-1},\sigma^{2})$, and 
consequently the measure-domination condition $\mathbb{P}_{X_k|Z_1^{k-1}}\ll \mathbb{Q}_{X_k|Z_1^{k-1}}$ holds.
\begin{align}
    \label{ekd1213}
    &D(\mathbb{P}_{X_k|Z_1^{k-1},A}\| \mathbb{Q}_{X_k|Z_1^{k-1}}|\mathbb{P}_{Z_1^{k-1},A})
 \\
  \nonumber
 &= \Epi{ \KL{\mathcal{N}(AX_{k-1}+U_{k-1},\sigma^{2})}{\mathcal{N}(0,\sigma^{2})} }
 \\
 \nonumber
 &= \frac{\log e}{2\sigma^2}\mathbb{E}_\mathbb{P}\left[(AX_{k-1}+U_{k-1})^2\right].
\end{align}
	The result follows by combining \eqref{newlowreddw} and~\eqref{ekd1213}.\hfill \,\QED
\section{Proof of Corollary~\ref{last2efthm1}}
Using \eqref{eq:plant} and \eqref{eq:watermaked_control}, we can rewrite $\oX_k$ and $X_k$ explicitly as follows
\begin{align*}
	\oX_k &= A^kX_0+\sum_{j=1}^{k-1}A^{k-1-j}(\oU_j+W_j) ,
 \\ X_k &= A^kX_0+\sum_{j=1}^{k-1}A^{k-1-j}(U_j+W_j) 
 \\ &= A^kX_0 + \sum_{j=1}^{k-1}A^{k-1-j}(\oU_j + \Gamma_j + W_j) 
\\\ &= \oX_k + \Psi_{k-1} \,.
\end{align*}
Thus, by~\eqref{eq:plant}, the following relation holds
\begin{align}\label{cmlwq02e2KNEWG}
AX_{k-1}+U_{k-1}=A\bar{X}_{k-1}+\bar{U}_{k-1}+\Psi_{k-1} \,.
\end{align}
	By comparing 
   \begin{align*}
   G(\bar{Z}_1^L) \triangleq \dfrac{\frac{\log e}{2\sigma^2}\sum_{k=1}^L \Epi{(A\bar{X}_{k-1}+\bar{U}_{k-1})^2}+1}{\log \left( R / \sqrt{\delta\beta}\right)},
   \end{align*}
   with 
  \begin{align*}
       G(Z_1^L) = \dfrac{\frac{\log e}{2\sigma^2}\sum_{k=1}^L \Epi{(A\bar{X}_{k-1}+\bar{U}_{k-1}+\Psi_{k-1})^2}+1}{\log \left( R / \sqrt{\delta\beta} \right)},
   \end{align*}
   in which we have utilized \eqref{cmlwq02e2KNEWG}, 
   and provided~\eqref{expt431mms}, we arrive at 
$G(\bar{Z}_1^L)>G(Z_1^L)$.\hfill \,\QED
\section{Proof of \lemref{lem:vecdetel}}
\label{lemaa345!3455}
 Since the hijacking phase of the vector analogue of the learning-based attack of~\eqref{mimicmodel} starts at time $k = L+1$, using~\eqref{eq:plant-vec} and~\eqref{eq:subtract4test}, we have
    \begin{subequations}
    \label{jf14053www_vec}
    \noeqref{jf14053www:basic,jf14053www:explicit,jf14053www:explicit_vec1}
    \begin{align}
    \nonumber
    &\sum_{k=1}^{T}  \left[ \textbf{y}_{k+1}-\texttt{A} \textbf{y}_k-\textbf{u}_k \right]   \left[ \textbf{y}_{k+1}-\texttt{A} \textbf{y}_k-\textbf{u}_k \right]^\dagger 
    \\
    \nonumber
    &=\sum_{k=1}^{L} \textbf{w}_k\textbf{w}_k^{\dagger}
    \\
    &\quad 
    + \sum_{k=L+1}^{T} \left(\btW_k+(\bhA-\texttt{A})\textbf{v}_k\right)\left(\btW_k+(\bhA-\texttt{A})\textbf{v}_k\right)^{\dagger} \quad\:\:
    \label{jf14053www:basic}
    \\
    &= \sum_{k=1}^{L}\textbf{w}_k\textbf{w}_k^{\dagger}+\sum_{k=L+1}^{T}\btW_k\btW_k^{\dagger}
    + \sum_{k=L+1}^{T} \left(\tbW_k \textbf{v}_k^{\dagger}(\bhA-\texttt{A})^{\dagger}\right)^{\dagger}
    \nonumber
    \\
    &\quad 
    + \sum_{k=L+1}^{T}
    (\bhA-\texttt{A})\textbf{v}_k \textbf{v}_k^{\dagger}(\bhA-\texttt{A})^{\dagger}
    +\sum_{k=L+1}^{T}\tbW_k \textbf{v}_k^{\dagger}(\bhA-\texttt{A})^{\dagger} \ \ 
    \label{jf14053www:explicit_vec}
    \end{align}
    \end{subequations}
    Let $\mathcal{F}_k$ be the $\sigma$-field 
    generated by $\{(\textbf{v}_k,\bhA,\textbf{w}_k,\textbf{u}_k)| k = L, \ldots, T - 1\}$. 
    Then, 
    $\textbf{v}_{k+1}^{\dagger}$ is $\mathcal{F}_k$ measurable, and $(\textbf{w}_{k+1},\mathcal{F}_k)$ is a martingale
    difference sequence, i.e., $\CE{\textbf{w}_{k+1}}{\cF_k} = 0$ a.s. 
    Consequently, using \cite[Lemma~2, Part~iii]{lai1982least}, we have
    
    \begin{align}
    \nonumber
    \\[-2\baselineskip]
    \begin{aligned}
        &\sum_{k=L+1}^{T} \tbW_k\textbf{v}_k^{\dagger} = O(1)
        \\
        &+ \begin{pmatrix} 
            o \left(\sum_{k=L+1}^{T} (\textbf{v}_k^{\dagger})_{1}^2 \right) & \ldots & o \left(\sum_{k=L+1}^{T} (\textbf{v}_k^{\dagger})_{n}^2 \right) \\
            \vdots & \vdots & \vdots
            \\
            o \left(\sum_{k=L+1}^{T} (\textbf{v}_k^{\dagger})_{1}^2 \right) & \ldots & o \left(\sum_{k=L+1}^{T} (\textbf{v}_k^{\dagger})_{n}^2 \right)
        \end{pmatrix}
        ~~~\mbox{a.s.}
    \end{aligned}
    \label{eflko1133-vec}
    \end{align}
    in the limit $T \to \infty$, 
    where  $(\textbf{v}_k^{\dagger})_{i}^2$ denotes the square of the $i$-th element of $\textbf{v}_k^{\dagger}$.
    Further, by the strong law of large numbers:
	\begin{align}
    \label{jjiief123331-vec}
	    \lim_{T \rightarrow \infty}\frac{1}{T} \left(\sum_{k=1}^{L}\textbf{w}_k^{\dagger}\textbf{w}_k+\sum_{k=L+1}^{T}\btW_k^{\dagger}\btW_k\right)=\bSigma \,~~\mbox{a.s.}
    \end{align}
     \indent 
     Substituting~\eqref{eflko1133-vec} 
    and \eqref{jjiief123331-vec} in \eqref{jf14053www:explicit_vec} completes the proof.\,\,\QED 


\section{Proof of Theorem~\ref{LB:dec-prob-vec}}
\eqref{k2e1jmfve!!!!1vec} follows from \lemref{lem:vecdetel} and 
\eqref{implici-vec3}.
%
We now prove~\eqref{k2e1jmfve!!!!2vec}. 
By the Law of total probability,  
\begin{align*}
    &\mathbb{P}_\texttt{A}\left(\|\bhA-\texttt{A}\|_{op}<\sqrt[]{\gamma\beta}\right) \ge  \mathbb{P}_\texttt{A}\left(\frac{1}{L} \sum_{k=1}^{L} \bX_k\bX_k^{\dagger}\succeq \zeta \texttt{I}_{n \times n}\right)
\\
&\qquad \cdot\mathbb{P}_\texttt{A}\left(\|\bhA-\texttt{A}\|_{op}<\sqrt[]{\gamma\beta} \middle| \frac{1}{L} \sum_{k=1}^{L} \bX_k\bX_k^{\dagger}\succeq \zeta \texttt{I}_{n \times n}\right).
\end{align*}
Since the control policy is $(\zeta,\rho)$-persistently exciting and $L-1 \ge L_0$, the 
result follows from \lemref{lem:Raginskyerror} and~\eqref{prexc345}.\hfill \,\QED 


\section{Proof of \lemref{lem:Raginskyerror}}
\label{lemaaapenx243333}

Since $\bG_{L-1}$ is a Hermitian matrix we start by noticing that since the event in~\eqref{prexc345} occurs for $L-1$ we have $\mbox{det}(\bG_{L-1}) \neq 0$, using
\cite[Theorem 7.8, part 2]{zhang2011matrix}. Thus, when the attacker  uses the LS estimation \eqref{learningAlgorithmvect-var} we deduce  
\begin{align*}
    \bhA-\texttt{A} &=\left(\sum_{k=1}^{L-1}\left((\bX_{k+1}-\bU_k)\bX^{\dagger}_k\right)-\texttt{A}\bG_{L-1}\right) \bG_{L-1}^{-1}
    \\
    &=\sum_{k=1}^{L-1}\left((\bX_{k+1}-\texttt{A}\bX_k-\bU_k)\bX_k^{\dagger}\right)\bG_{L-1}^{-1}
    \\
    &=\sum_{k=1}^{L-1}\left(\bW_k\bX_k^{\dagger}\right)\bG_{L-1}^{-1},
\end{align*}
where the last two equalities  follow from~\eqref{Gram_matrix_1} and~\eqref{eq:plant-vec}, respectively. Thus, using sub-multiplicativity of operator nroms and the triangle inequality we have
\begin{align}
\label{eq:norm1rmrg}
    \|\bhA-\texttt{A}\|_{op} \le  \sum_{k=1}^{L-1}\|\textbf{w}_k\textbf{x}_k^{\dagger}\|_{op} \|\bG_{L-1}^{-1}\|_{op}.
\end{align}
We now continue by upper bounding $\|\bG_{L-1}^{-1}\|_{op}$ as follows.
Since the event in~\eqref{prexc345} occurs for $L-1$, using
\cite[Theorem 7.8, part 3]{zhang2011matrix} we deduce
\begin{align}
\label{eq:355211!4}
    \frac{1}{\zeta L}\upsilon^{\dagger}\Id_{n \times n}\upsilon \ge \upsilon^{\dagger}\bG_L^{-1}\upsilon 
\end{align}
for all $\upsilon \in \real^{n \times 1}$. Since $\bG_{L-1}$ is a Hermitian positive semidefinite matrix, then so is $\bG_{L-1}^{-1}$ (see \cite[Problem 1, Section 7.1]{zhang2011matrix}). Thus, using \cite[Theorem 7.4]{zhang2011matrix} the Hermitian matrix $\sqrt{\bG_{L-1}^{-1}}$ exists. We continue by noticing $\upsilon^{\dagger}\Id_{n \times n}\upsilon=\|\upsilon\|^2$, and
\begin{align*}
    \upsilon^{\dagger}\sqrt{\bG_{L-1}^{-1}}^{\dagger}\sqrt{\bG_{L-1}^{-1}}\upsilon=\left\|\sqrt{\bG_{L-1}^{-1}}\upsilon\right\|^2.
\end{align*}
Thus, using~\eqref{eq:355211!4} we deduce
\begin{align}
\label{eq:opt344sqrt}
  \left\|\sqrt{\bG_{L-1}^{-1}}\right\|_{op} \le \frac{1}{\sqrt{\zeta L}}.
\end{align}
Using~\eqref{eq:opt344sqrt}, sub-multiplicativity of operator nroms, and~\eqref{eq:norm1rmrg},~\eqref{upperlearning223-first} follows.\hfill \,\QED 
\section{Proof of Theorem~\ref{LB:dec-prob-nonlinear3}}
\label{sectuio3jtgjo403222}
We start by proving the following lemma, which is an extension of the \lemref{Ineedtowk} to our nonlinear setup, and it shows that the deception probability is related to the performance of the learning algorithm. 
\begin{lemma}
\label{nonlineaIneedtowkq}
Given $T = L + c$, consider any learning-based attack~\eqref{mimicmodelnonoe33} and some measurable control policy $\{U_k\}$.
    If  
    \begin{align*}
     &\frac{1}{L+c} \sum_{k=L+1}^{L+c}[f(V_k, U_k)-\hat{F} (V_k, U_k)]^2+
     \\
     &\frac{2}{L+c} \sum_{k=L+1}^{L+c} |\tW_k| |\hat{F} (V_k, U_k)-f(V_k, U_k)| 
    \le \delta
    \end{align*}
    holds in the limit of $L \to \infty$, then the attacker is able to deceive the controller and remain undetected  a.s.
\end{lemma}
\textit{Proof of \lemref{nonlineaIneedtowkq}}:
Using~\eqref{mimicmodelnonoe33} we have
\begin{align*}
    V_{k+1}-f(V_k,U_k)=\tW_k+\hat{F} (V_k, U_k)-f(V_k, U_k).
\end{align*}
Thus, given the test~\eqref{test4nonlinear443}, the attacker manages to deceive the controller and remain undetected 
if 
\begin{align*}
\begin{aligned}
\frac{1}{T} \left(\sum_{k=1}^{L} W_k^2+ \sum_{k=L+1}^{T} (\tW_k+\hat{F} (V_k, U_k)-f(V_k, U_k))^2\right)
 \\  \in (\Var{W}-\delta, \Var{W}+\delta).
\end{aligned}
\end{align*}
We have
\begin{align}
\nonumber
    &\frac{1}{T} \left(\sum_{k=1}^{L} W_k^2+ \sum_{k=L+1}^{T} (\tW_k+\hat{F} (V_k, U_k)-f(V_k, U_k))^2\right)
    \\
    \nonumber
    &= \frac{1}{T} \left(\sum_{k=1}^{L} W_k^2+\sum_{k=L+1}^{T}\tW_k^2\right)
    \\
    \nonumber
    &\quad + \frac{1}{T} \sum_{k=L+1}^{T}(\hat{F} (V_k, U_k)-f(V_k, U_k))^2
    \\
  \label{nonlineat4553}
    &\quad +\frac{2}{T} \sum_{k=L+1}^{T} \tW_k (\hat{F} (V_k, U_k)-f(V_k, U_k)) \,. \qquad
\end{align}
Since $T=L+c$, by taking $L$ to infinity and substituting \eqref{jjiief123331} in \eqref{nonlineat4553} we deduce if
    \begin{align*}
     &\frac{1}{L+c} \sum_{k=L+1}^{L+c}(f(V_k, U_k)-\hat{F} (V_k, U_k))^2+
     \\
     &\frac{2}{L+c} \sum_{k=L+1}^{L+c} \tW_k (\hat{F} (V_k, U_k)-f(V_k, U_k)) \in
    (-\delta,\delta).
    \end{align*}
    holds in the limit of~$L \to \infty$, then the attacker remains undetected  a.s. 	The result follows using triangle inequality.~\oprocend
    
We continue by re-stating the following lemma from
\cite{chowdhury2017kernelized,berkenkamp2017safe}  for our setup.     
\begin{lemma}
\label{felixlemaakerme33}
 Consider the nonlinear plant~\eqref{nonlinearsystem1}, where $\|f\|_\mathcal{K} \le \chi$. 
   If
   the attacker chooses \rm{$\hat{F}=\mathcal{M}_{L-1}$} according to the GP-based inference \eqref{GPalgoriti345666},
    then
    \begin{align*}
        \mathbb{P}_f\left(|f(V_k,U_k)-\hat{F}(V_k,U_k)| \le \varrho_k \sigma_{L-1} (V_k,U_k)\right) \ge 1-\xi_k,
    \end{align*}
for $k \ge L+1$, where $\varrho_k \triangleq \chi+4\sigma \sqrt{\psi_{L-1}+1+\ln(\frac{1}{\xi_k})}$, and $\psi_{L-1}$ is defined as follows
\begin{align*}
       \psi_{L-1} &\triangleq \frac{1}{2}\sum_{k=1}^{L-1}\ln(1+\sigma^{-2}\sigma_{k-1}^2(Z_k)).
\end{align*}
\end{lemma}

Define $\nu_k \triangleq \varrho_k \sigma_{L-1} (V_k,U_k)$, then $\xi_k$ can be calculated as~\eqref{akhr349595922!!}. Also, using \lemref{felixlemaakerme33}, we deduce
\begin{align*}
      &\frac{1}{L+c} \sum_{k=L+1}^{L+c}[\hat{F} (V_k, U_k)-f(V_k, U_k)]^2+
      \\
      &\frac{2}{L+c} \sum_{k=L+1}^{L+c} |\tW_k| |\hat{F} (V_k, U_k)-f(V_k, U_k)| 
       \le
       \\
       &\frac{1}{L+c}\sum_{k=L+1}^{L+c} \nu_k^2 + \frac{2}{L+c} \sum_{k=L+1}^{L+c} |\tW_k|\nu_k 
\end{align*}
holds at least with probability $\prod_{k=L+1}^{L+c} (1-\xi_k)$.
The result now follows using~\lemref{nonlineaIneedtowkq}, and~\eqref{iefi54076543!!!1}.\hfill \,\QED 
\end{document}